\documentclass[final,3p,times,authoryear,]{elsarticle}

\makeatletter
\def\ps@pprintTitle{%
    \let\@oddhead\@empty
    \let\@evenhead\@empty
}
\makeatother

\usepackage{amssymb}
\usepackage{amsmath}
\usepackage{amsthm}

\usepackage{booktabs}

\usepackage{setspace}
\usepackage{lineno}
\usepackage{bm}
\usepackage[hidelinks,colorlinks=false]{hyperref}
\usepackage{xcolor}
\definecolor{brgreen}{HTML}{004225}
\definecolor{avio}{HTML}{5D8AA8}

\usepackage[font={footnotesize},labelfont=bf]{caption,subfig}
\usepackage{siunitx}
\DeclareSIUnit\bar{bar}
\DeclareSIUnit\pixels{pixels}
\DeclareSIUnit\pixel{pixel}
\DeclareSIUnit\fps{FPS}
\newcommand{\ie}{i.\,e.~}

\begin{document}

\begin{frontmatter}



\title{
Dynamic morphing of non-Euclidean active shells \\[1.5mm] enabled by flutter under a steady stimulus
} %% Article title


\author[SISSA]{{\large Roberto Marchello}}
\author[SISSA]{{\large Pietro Maria Santucci}}
\author[SISSA]{{\large Giovanni Noselli}}
\affiliation[SISSA]{
            organization={{\noexpand\normalsize
        SISSA\,--\,International School for Advanced Studies}},
            addressline={{\noexpand\normalsize Via Bonomea 265}}, 
            city={{\noexpand\normalsize Trieste}},
            postcode={{\noexpand\normalsize 34136}}, 
            country={{\noexpand\normalsize Italy}}
}

%% Abstract
\begin{abstract}

\noindent Synthetic active surfaces are usually designed to attain a prescribed static configuration, with dynamic shape changes typically requiring modulation of the actuation.
We show that thin active shells can instead sustain dynamic morphing under a steady stimulus by harnessing flutter instability. 
We focus on polyelectrolyte hydrogel shells immersed in a viscous fluid and subjected to a constant and uniform electric field, and specialize the non-Euclidean framework for thin structures to account for field-induced activity through a spontaneous curvature evolving in time with the system configuration.
Examining elliptic plates and cylindrical shells as test cases, we integrate linear stability analyses with nonlinear simulations to first locate the onset of flutter and then unveil the resulting dynamics, characterized by thrust-generating shape changes. 
In all cases, the emergent behavior stems from the interplay between active reconfiguration and elasticity in the fluid environment.
For plates, the problem is purely flexural in the linear regime, while the competition between bending and stretching becomes crucial in selecting the ensuing shapes and swimming trajectories at large deformations. 
Conversely, for cylindrical shells bending and stretching are coupled at all levels, with the latter becoming essential to sustain flutter.
Experiments on millimeter-sized hydrogel plates provide tangible support for our theoretical findings through shape reconstructions performed by combining digital image correlation with optical stereo microscopy.
Flutter thus offers a route to untethered soft robotic and microfluidic devices based on active shells and operating under a stimulus carrying neither an intrinsic frequency nor a characteristic length scale.
\end{abstract}

%% Keywords

\end{frontmatter}

%\linenumbers


%\comm{Page height: \the\textheight, Page width: \the\textwidth}

%\comm{Fontsize: testo 10pt, caption 8pt}

%%%%%%%%%%%%%%%%%%%%%%%%%%%%%%%%%%%%%%%%%%%%%%%%%%%%%%%%%%%%%%%%

\section{Introduction}
\label{sec:intro}

Many biological structures through which living organisms move, feed and transport fluid take the form of thin shells. The leaves of a Venus flytrap close on its prey within a fraction of a second by snapping between two curved configurations~\citep{forterreHowVenusFlytrap2005}; the bell of a swimming medusa deforms periodically to propel itself~\citep{costello2021hydrodynamics}; the wall of a tubular organ transports its contents by a traveling contraction~\citep{shapiroJaffrinWeinberg1969}.
In each of these cases, motion is driven by internal activity rather than by external mechanical forcing: the material dynamically modifies its local, stress-free configuration and the structure  deforms in response.

Reconfigurations of thin active shells are constrained by \emph{Gauss's Theorema Egregium}~\citep{IntroductionDifferentialGeometry2005}.  
Specifically, when the metric cannot accommodate for the prescribed curvature, no configuration exists relieving the internal stresses entirely: the body is geometrically frustrated, and the shape it takes is a compromise between bending and stretching energies.
This is the mechanics of \emph{non-Euclidean} morphing~\citep{efratiElasticTheoryUnconstrained2009}, which underlies shape changes as diverse as the opening of chiral seed pods, the rippling of long leaves and the buckling of responsive gel sheets~\citep{klein_efrati_sharon_2007,  armonGeometryMechanicsOpening2011, liangMahadevan2009, kimHayward2012}.

At present, synthetic active surfaces have mostly been engineered to morph and settle into a target shape~\citep{pezzullaCurvaturedrivenMorphingNonEuclidean2017,pezzullaCurvatureInducedInstabilitiesShells2018,mahadevan_2016}.
Sustained morphing in slender structures has been achieved only recently, and primarily through complex actuation strategies: by activating the material in space and time along a prescribed path~\citep{biggins_2025}, by modulating an external field, as in magnetically actuated artificial cilia and swimmers~\citep{dreyfus2005microscopic, Miao2021, zhou2021magnetically} or in light-driven liquid crystal elastomers~\citep{sartoriPhotochemicallyInducedPropulsion2024},
or by embedding in the synthetic structure a chemical oscillator~\citep{levin2020}.
An intriguing alternative is to exploit the intrinsic mechanics of the structure and its interactions with the environment to allow a target sustained motion to emerge from a stimulus that is constant in time and uniform in space.
In this respect, recent experiments have shown that filaments of polyelectrolyte hydrogel (PEH) may undergo nonreciprocal, large-amplitude oscillations when exposed to a steady and uniform electric field~\citep{cicconofriNonreciprocalOscillationsPolyelectrolyte2023, boiardiMinimalActuationControl2024}.
These deformations have been shown to originate from flutter~\citep{bigoniFlutterInstabilitySolids2023}, whereby the loss of stability of an equilibrium gives rise to oscillations of growing amplitude, which eventually settle into a limit cycle, as in a Hopf bifurcation. Being nonreciprocal, such periodic motions escape the constraint that Purcell's scallop theorem~\citep{purcell_1977} places on locomotion at low Reynolds number, thus allowing a filament to generate a net thrust or to swim in a viscous environment, with a periodic beat that mimics the dynamics of eukaryotic cilia and flagella.

In active filaments, curvature-driven reconfigurations are not subjected to geometric frustration. 
In this work, we investigate whether the flutter-based actuation mechanism identified for PEH filaments~\citep{cicconofriNonreciprocalOscillationsPolyelectrolyte2023, boiardiMinimalActuationControl2024} persists in the context of thin active shells, and how geometric frustration influences its onset and subsequent dynamics.
To this end, we develop a nonlinear model for thin PEH shells that actively deform in response to external electric stimuli and are free to move in a viscous environment at low Reynolds numbers.
We specialize the model to two complementary geometries:  elliptic plates and cylindrical shells.
The first case study extends the ribbon-like geometry considered in~\citet{boiardiMinimalActuationControl2024} to two dimensions. At small aspect ratio, an elliptic plate approaches the previously explored slender-strip limit, while it  reveals genuinely two-dimensional effects as the aspect ratio grows.
The second case study, by contrast, has no one-dimensional counterpart. A cylindrical shell with circular cross-section provides the simplest, intrinsically curved geometry of practical interest, as tubular soft actuators are used in microfluidics to transport fluid by peristalsis~\citep{heElectricallyControlledLiquid2019, najiyaContinuousFlowMicrofluidic2023}. We study the active response of a cylindrical shell  to electric stimuli applied along its axis.

For both geometries, the analysis is developed along the same line.
We identify a trivial equilibrium configuration and linearize the governing equations about it to asses the stability of the system by solving the resulting eigenvalue problem. In particular, we identify the regions of the parameter space where the system undergoes flutter and the modes that drive the instability.
The behavior of the two geometries differ in an essential respect: while the linearized problem for a plate is insensitive to stretching, we prove that an unstretchable cylindrical shell cannot flutter, regardless of the electric field strength, so that stretching is necessary to sustain the instability.
We then solve numerically the nonlinear equations in time, to investigate the dynamics of the system beyond the onset of the instability, typically leading to a limit cycle.
For elliptic plates, we find that the large-amplitude shapes differ from those predicted by the linear stability analysis: they exhibit nearly vanishing Gaussian curvature and break the symmetry with respect to the direction of motion. For cylindrical shells, we find instead that the linear unstable
modes are predictive of the dynamics emerging at large displacements.

We complement the analysis with experiments on millimeter-sized PEH elliptic plates, fabricated with the aspect ratio used in the numerical simulations.
Above a threshold value of the applied voltage, we observe the onset of flutter instability and the resulting motion of the plate, thus extending to a genuinely two-dimensional structure the behavior so far reported only for PEH filaments and ribbons~\citep{cicconofriNonreciprocalOscillationsPolyelectrolyte2023, boiardiMinimalActuationControl2024}.
The three-dimensional shape of the plate is reconstructed in time by stereo digital image correlation~\citep{ Sutton2009-DIC}, a technique previously applied, together with optical stereo microscopy and fluorescence-based speckle, to small-scale biological and bioinspired samples in static conditions~\citep{Schreier2004, Zhang2006, Sutton2008-JBMRPA, Berfield2006, Hu2013, Hu2016}.
%to the authors' knowledge, the shape and the motion of a freely swimming active shell have not been reconstructed before.
%
The measurements resolve the retrograde traveling wave that propels the plate and the ruled-surface character of the deformed shapes, in quantitative agreement with the nonlinear simulations.

The paper is organized as follows: Section~\ref{sec:nonlinearModel} develops the nonlinear model for active non-Euclidean shells, Sections~\ref{sec:plates} and~\ref{sec:cylindricalShells} specialize it to elliptic plates and to cylindrical shells, Section~\ref{sec:experiments} reports the experiments on plates and their comparison with the numerical predictions, and Section~\ref{sec:conclusions} draws the conclusions.

%%%%%%%%%%%%%%%%%%%%%%%%%%%%%%%%%%%%%%%%%%%%%%%%%%%%%%%%%%%%%%%%

\section{A nonlinear model for non-Euclidean active shells}
\label{sec:nonlinearModel}

In this section, we develop a nonlinear mechanical model for an active thin PEH shell that captures its dynamics in response to an external electric stimulus.
Since the shell thickness is much smaller than the characteristic dimensions of the mid-surface, we adopt Koiter's nonlinear theory for unshearable shells~\citep{koiter1966nonlinear,IntroductionDifferentialGeometry2005}, whose consistent first approximation~\citep{koiterw.t.ConsistentFirstApproximation,steigmann2013koiter} reduces the three-dimensional problem to a two-dimensional one for the mid-surface alone.
To account for activity, the Koiter model is recast within the framework of \emph{non-Euclidean shell theory}~\citep{efratiElasticTheoryUnconstrained2009,efiefratieransharonrazkupfermanNonEuclideanPlatesShells}, which describes shells that lack a stress-free rest state because of a geometric incompatibility.
The incompatibility originates from the internal activity of the material, which generates a spherical \emph{spontaneous curvature}, such that the shell becomes frustrated under the external stimulus.
This is the mechanism underlying shape morphing in thin biological and synthetic structures~\citep{klein_efrati_sharon_2007,armonGeometryMechanicsOpening2011,liangMahadevan2009,kimHayward2012,mahadevan_2016}. 
Similar modeling approaches have been used for bilayer shells~\citep{pezzullaCurvaturedrivenMorphingNonEuclidean2017,pezzullaCurvatureInducedInstabilitiesShells2018}, for curvature-driven instabilities in thin active
shells~\citep{giudiciCurvaturedrivenInstabilitiesThin2022}, and for chemo-mechanical theories of
active surfaces and of the cell cortex~\citep{salbreuxMechanicsActiveSurfaces2017,yinBiochemomechanicalTheoryActive2021,yinThreedimensionalChiralMorphodynamics2022},
while a non-Euclidean shell model accounting for large strains is developed
by~\citet{taffetaniNonlinearMorphoelasticEnergy2026}.
In contrast to many of these works, where the natural fundamental forms are prescribed a priori, here the spontaneous curvature evolves in time and its evolution is coupled to the current configuration of the shell.

\subsection{Geometry and kinematics}
\label{sec:nonlinearModel:geometry}

\begin{figure}
    \centering
    \includegraphics[width=\linewidth]{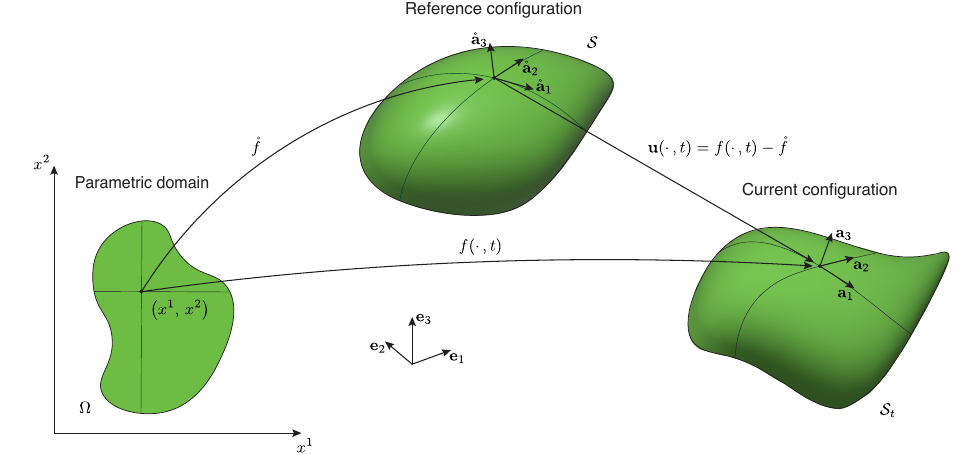}
    \caption{Parametric domain $\Omega$, reference configuration $\mathcal{S}$, and current configuration $\mathcal{S}_t$ of the shell's mid-surface.
    The map $\mathring{f}$ identifies the reference configuration $\mathcal{S}$, on which the basis vectors $\mathring{\mathbf{a}}_i$ are defined.
    After deformation, the map $f(\cdot \, , t)$ identifies the current configuration $\mathcal{S}_t$ at time $t \geq 0$, with associated basis vectors $\mathbf{a}_i$.
    The displacement vector between $\mathcal{S}$ and $\mathcal{S}_t$ is defined as $\mathbf{u}(\cdot \, ,t) = f(\cdot \, ,t) - \mathring{f}$.
    }
    \label{fig:model}
\end{figure}

We follow the presentation of~\citet{IntroductionDifferentialGeometry2005} to briefly review the fundamental notions on the geometry of surfaces relevant to our study.
Throughout the manuscript Latin indices range in $\{1,2,3\}$, and throughout this section Greek indices range in $\{1,2\}$.
We denote by $\mathcal{E}$ the three-dimensional Euclidean point space and by $\mathcal{V}$ the associated vector space of translations, equipped with a fixed Cartesian coordinate frame $\{ \mathbf{e}_1, \mathbf{e}_2, \mathbf{e}_3\}$ with origin $o \in \mathcal{E}$.
Let $\Omega \subset \mathbb{R}^2$ denote the \emph{parametric domain} for the curvilinear coordinates $x^{\alpha}$.
The immersion  $\mathring{f}:\Omega \to \mathcal{E}$ identifies the \emph{reference configuration} of the shell's mid-surface, $\mathcal{S}=\mathring{f} (\Omega)$, see Fig.~\ref{fig:model}.
The covariant basis $\mathring{\mathbf{a}}_i(x^{\alpha})$ is defined by the vectors $\mathring{\mathbf{a}}_\alpha = \partial_\alpha \mathring{f}$ tangent to $\mathcal{S}$ in $\mathring{f}(x^\alpha)$ and by the unit normal $\mathring{\mathbf{a}}_3 = \mathring{\mathbf{a}}_1 \times \mathring{\mathbf{a}}_2/|| \mathring{\mathbf{a}}_1 \times \mathring{\mathbf{a}}_2||$, where $|| \cdot ||$ denotes the Euclidean norm.
In parallel, the vectors $\mathring{\mathbf{a}}^i(x^\alpha)$ defined as $\mathring{\mathbf{a}}^i \cdot \mathring{\mathbf{a}}_j = \delta^i_{j}$ form the contravariant basis.
We recall that $\mathring{\mathbf{a}}^3 = \mathring{\mathbf{a}}_3$ and that $\mathring{\mathbf{a}}^\beta$ are also tangent to $\mathcal{S}$ in $\mathring{f}(x^\alpha)$.
Unless specified, we will implicitly assume the dependence on the coordinates $x^\alpha$.

The covariant and contravariant components
%$\mathring{a}_{\alpha \beta}$ and $\mathring{a}^{\alpha \beta}$
of the first fundamental form or \emph{metric tensor} $\mathring{\mathbf{a}}$ of $\mathcal{S}$ are $\mathring{a}_{\alpha \beta} = \mathring{\mathbf{a}}_\alpha \cdot \mathring{\mathbf{a}}_\beta$ and $\mathring{a}^{\alpha \beta} = \mathring{\mathbf{a}}^\alpha \cdot \mathring{\mathbf{a}}^\beta$, respectively, where $\cdot$ denotes the Euclidean scalar product. These encode variations of lengths and angles on the reference surface.
Likewise, the covariant and mixed components of the second fundamental form or \emph{curvature tensor} $\mathring{\mathbf{b}}$ of $\mathcal{S}$ are $\mathring{b}_{\alpha \beta} = \mathring{\mathbf{a}}_3 \cdot \partial_\beta \mathring{\mathbf{a}}_\alpha$ and $\mathring{b}_{\alpha}^{\beta} = \mathring{a}^{\beta \sigma} \mathring{b}_{\sigma \alpha}$, respectively, where Einstein's summation for repeated indices is employed. These  describe changes in the orientation or rotation of the normal to the  surface.
We highlight that the tensors $\mathring{\mathbf{a}}$ and $\mathring{\mathbf{b}}$ satisfy Gauss--Codazzi--Mainardi compatibility equations, since they are derived from an immersion~\citep{IntroductionDifferentialGeometry2005}.

After deformation, we denote by $\mathcal{S}_t$ the \emph{current configuration} of the shell's mid-surface at time $t \geq 0$, identified by the immersion $f : \Omega \times [0, +\infty) \to \mathcal{E}$ such that $\mathcal{S}_t = f(\Omega,t)$.
We define the \emph{displacement vector field} $\mathbf{u} : \Omega \times [0, +\infty) \to \mathcal{V}$ as $\mathbf{u}(\cdot \, ,t) = f(\cdot \, ,t) - \mathring{f}$, see Fig.~\ref{fig:model}.
The vectors that form the covariant basis of $\mathcal{S}_t$ are $\mathbf{a}_\alpha = \partial_\alpha f$ and $\mathbf{a}_3 = \mathbf{a}_1 \times \mathbf{a}_2 / || \mathbf{a}_1 \times \mathbf{a}_2 ||$, while the covariant components of the fundamental forms of $\mathcal{S}_t$ are $a_{\alpha \beta} = \mathbf{a}_\alpha \cdot \mathbf{a}_\beta$ and $b_{\alpha \beta} = \mathbf{a}_3 \cdot \partial_\beta \mathbf{a}_\alpha$.
The velocity of the current shell's mid-surface is denoted by $\partial_t \mathbf{u}$.

\subsection{Elastic energy}
\label{sec:nonlinearModel:energy}

In standard compatible elasticity, the reference metric tensor $\mathring{\mathbf{a}}$ and the reference curvature tensor $\mathring{\mathbf{b}}$ identify the stress-free state of the shell and define the base state with respect to which strains are measured.
In contrast, active materials subjected to external stimuli, such as light, temperature variations, or electro-magnetic fields~\citep{sartoriPhotochemicallyInducedPropulsion2024, chenStimuliresponsiveHydrogelsPromising2024}, may develop complex configurations with residual stresses in the absence of external loading, as a consequence of internal processes~\citep{klein_efrati_sharon_2007,levin2020}.
As a result, such materials generally possess no stress-free reference state, and elastic strains must instead be measured relative to a \emph{stress-free natural configuration}, which in general cannot be embedded in the three-dimensional Euclidean space and depends on the specific stimulus applied.

In the context of active shells, pioneering works~\citep{efratiElasticTheoryUnconstrained2009,armonGeometryMechanicsOpening2011} have paved the way for a rigorous mathematical and mechanical treatment of large deformations in two-dimensional inelastic structures in the small strain regime, termed \emph{non-Euclidean shells}.
In this framework, the stress-free natural state is characterized by the \emph{natural metric tensor} $\bar{\mathbf{a}}$ and the \emph{natural curvature tensor} $\bar{\mathbf{b}}$.
While $\mathring{\mathbf{a}}$ and $\mathring{\mathbf{b}}$ are derived from the map $\mathring{f}$, the natural tensors $\bar{\mathbf{a}}$ and $\bar{\mathbf{b}}$ must be prescribed according to the applied stimulus and may be incompatible, in the sense that they do not generally satisfy the Gauss--Codazzi--Mainardi compatibility equations.
Consequently, they do not correspond to the metric and curvature tensors of a surface embedded in the three-dimensional Euclidean space.

Under the hypothesis of small strains, plane-stress conditions, and reformulating standard Kirchhoff--Love assumptions~\citep{love1888h}, \citet{efratiElasticTheoryUnconstrained2009} derived the reduced two-dimensional elastic energy of a non-Euclidean plate, later generalized to non-Euclidean shells~\citep{efiefratieransharonrazkupfermanNonEuclideanPlatesShells,armonGeometryMechanicsOpening2011}.
The resulting elastic energy reads
\begin{equation}
\label{eq:shellElasticEnergy}
	U = \frac{1}{2} \int_\Omega \left( N^{\alpha \beta}  G_{\alpha \beta} + M^{\alpha \beta} R_{\alpha \beta}  \right) \sqrt{\bar{a}} \, \mathrm{d}x^1 \mathrm{d}x^2 ,
\end{equation}
where $\bar{a} := \mathrm{det} ( \bar{\mathbf{a}} )$ and 
\begin{equation}
\label{eq:NonLinForcesMoments}
	N^{\alpha \beta} = h \mathcal{A}^{\alpha \beta \gamma \delta} G_{\gamma \delta} , 
	\qquad M^{\alpha \beta} = \frac{h^3}{12} \mathcal{A}^{\alpha \beta \gamma \delta} R_{\gamma \delta} 
\end{equation}
are the contravariant components of the \emph{membrane internal forces} and \emph{internal moments}, respectively, $h$ is the shell thickness, and
\begin{equation}
\label{eq:changeMetricCurvature}
	G_{\alpha \beta} = \frac{1}{2} \left( a_{\alpha \beta} - \bar{a}_{\alpha \beta} \right) ,
	\qquad R_{\alpha \beta} = b_{\alpha \beta} - \bar{b}_{\alpha \beta} 
\end{equation}
are the covariant components of the \emph{change of metric} and \emph{change of curvature tensors} with respect to the natural state, respectively.
The contravariant components of the fourth-order reduced elasticity tensor are
\begin{equation}
	\mathcal{A}^{\alpha \beta \gamma \delta} = \frac{Y}{1+\nu} \left( \frac{\nu}{1-\nu} \bar{a}^{\alpha \beta} \bar{a}^{\gamma \delta} + \bar{a}^{\alpha \gamma} \bar{a}^{\beta \delta} \right) ,
\end{equation}
where $Y$ and $\nu$ denote the \emph{Young modulus} and the \emph{Poisson ratio} of the material, respectively.
As standard in shell theories, the elastic energy \eqref{eq:shellElasticEnergy} is the sum of a \emph{membrane energy}, proportional to $h$, and a \emph{bending energy}, proportional to $h^3$. Consequently, for sufficiently small thickness, stretching is energetically more costly than bending. 

It can be shown~\citep{ciarletJUSTIFICATIONTWODIMENSIONALNONLINEAR2001,koiter1966nonlinear} that replacing the change of curvature $R_{\alpha \beta}$ in~\eqref{eq:changeMetricCurvature}$_2$ with
\begin{equation}
\label{eq:simplCurvaTensor}
	\sqrt{\frac{a}{\bar{a}}} \, b_{\alpha \beta} - \bar{b}_{\alpha \beta} ,
\end{equation}
where $a := \mathrm{det} ( \mathbf{a} )$, introduces an error of the same order as the global approximation error of the reduced model. The two formulations are therefore energetically equivalent. This alternative expression is particularly convenient, as it avoids potential divisions by zero and simplifies the explicit form of the energy; it will therefore be adopted throughout the remainder of the paper.

\subsection{Active response of a PEH shell to an electric field}
\label{sec:nonlinearModel:activity}

\begin{figure}
    \centering
    \includegraphics[width=1\linewidth]{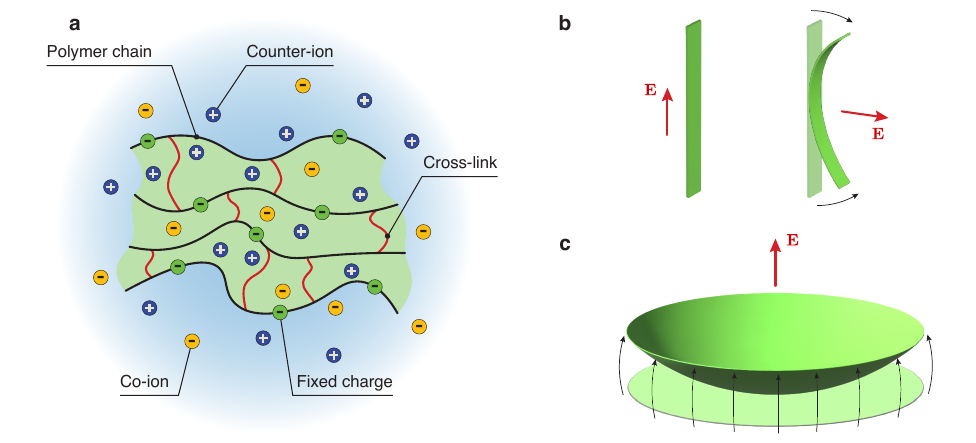}
    \caption{
    Active response of a polyelectrolyte hydrogel (PEH) to an external electric field.
    (a) Schematic of a PEH, consisting of an ionic polymer network with fixed charges and cross-links, swollen in an electrolytic solution containing mobile counter-ions and co-ions.
    (b) A PEH ribbon subjected to an electric field: no deformation occurs when the field is parallel to the filament axis (left), whereas a field applied across the thickness induces bending due to differential swelling (right).
    (c) Analogously, a PEH shell subjected to an electric field along the surface normal develops a locally spherical curvature, since no in-plane direction is preferred.
    }
    \label{fig:GelActivity}
\end{figure}

As mentioned in Section \ref{sec:nonlinearModel:energy}, the tensors $\bar{\mathbf{a}}$ and $\bar{\mathbf{b}}$ of natural metric and curvature must be formulated ad hoc in relation to the specific external stimulus under consideration.
Here, we specialize the model to capture the active response of thin shells made of PEH subjected to an external electric field $\mathbf{E}$ while immersed in an ionic solution.
The proposed framework extends and complements the models proposed by~\citet{cicconofriNonreciprocalOscillationsPolyelectrolyte2023},~\citet{boiardiMinimalActuationControl2024},~\citet{boiardiBreakingLeftrightSymmetry2024}, and~\citet{boiardi2026three} in the context of active filaments.
PEHs respond to electric fields by undergoing large deformations, an active behavior that originates from the ionic nature of their polymer network and the presence of mobile ionic species in the fluid phase~\citep{hongLargeDeformationElectrochemistry2010,narayan2022coupled}, see Fig.~\ref{fig:GelActivity}a.
When an electric field is applied across the thickness of a PEH filament, differential swelling induced by ionic redistribution leads to bending~\citep{boiardiMinimalActuationControl2024,liSpontaneousRapidElectroactuated2022}, as illustrated in Fig.~\ref{fig:GelActivity}b.
Owing to their elongated geometry, ribbons develop curvature primarily in a single direction.
By contrast, for PEH shells no in-plane direction is preferred.
As a result, differential swelling induced by an electric field applied along the shell normal is expected to generate a locally spherical curvature, with negligible changes in thickness, see Fig.~\ref{fig:GelActivity}c.
However, for thin shells, the formation of large spherical curvatures is energetically costly, due to the strong coupling between bending and stretching.
We therefore anticipate a competition between active processes and elastic energies, leading to complex, dynamic shape changes in PEH shells, as we reveal in Sections~\ref{sec:plates} and~\ref{sec:cylindricalShells}.

We encode this active behavior through a single scalar field, the \emph{spontaneous spherical curvature} $\kappa : \Omega \times [0,+\infty) \to \mathbb{R}$, whose evolution in time is specified at the end of this section.
More specifically, we assume differential swelling to be in-plane isotropic and to vary linearly across the shell thickness, while vanishing at the mid-surface. 
Hence, the natural metric coincides with the reference one, while the swelling gradient across the thickness induces spontaneous bending, such that the natural tensors are
\begin{equation}
\label{eq:refTensors}
	\bar{a}_{\alpha\beta} = \mathring{a}_{\alpha\beta} , \qquad
	\bar{b}_{\alpha\beta} = \mathring{b}_{\alpha\beta} - \kappa \, \mathring{a}_{\alpha\beta} ,
\end{equation}
with a negative $\kappa$ corresponding to bending of the shell in the direction of its normal \citep{marchelloPhDThesis2026}.
Since $\bar{a} = \mathring{a}$, where $\mathring{a} := \mathrm{det} ( \mathring{\mathbf{a}})$, the natural and reference area elements coincide, and the two metrics also coincide in raising and lowering indices.
We remark that the same natural tensors were adopted, on a different physical basis, for the bilayer shells studied in~\citet{pezzullaCurvaturedrivenMorphingNonEuclidean2017}.
Finally, gathering~\eqref{eq:changeMetricCurvature}, \eqref{eq:simplCurvaTensor} and~\eqref{eq:refTensors}, the covariant components of the change of metric and curvature tensors with respect to the natural state are
\begin{equation}
\label{eq:changeMetricCurvature2}
	G_{\alpha \beta} = \frac{1}{2} \left( a_{\alpha \beta} - \mathring{a}_{\alpha \beta} \right) ,
	\qquad 
    R_{\alpha \beta} = \sqrt{\frac{a}{\mathring{a}}} \, b_{\alpha \beta} - \mathring{b}_{\alpha \beta} + \kappa \, \mathring{a}_{\alpha \beta} .
\end{equation}
Notice that in the undeformed reference state $G_{\alpha \beta} = 0$, whereas $R_{\alpha \beta} = \kappa \, \mathring{a}_{\alpha \beta}$, so that the shell is frustrated for $\kappa \neq 0$ and carries a residual bending strain that cannot be entirely relieved without stretching, as dictated by Gauss's Theorema Egregium.

It remains to specify how the spontaneous curvature $\kappa$ evolves in time in relation to the external driving stimulus. Here we adopt the evolution law
\begin{equation}
\label{eq:evLaw}
	\tau \partial_t \kappa + \kappa = - \kappa_E \, \mathbf{e} \cdot \mathbf{a}_3 ,
\end{equation}
where $\tau$ is a characteristic relaxation time, $\kappa_E$ is a target curvature, and $\mathbf{e} =  \mathbf{E} / ||\mathbf{E}||$ is the unit vector encoding the direction of the electric field.
This equation is reminiscent of the evolution laws proposed by~\citet{cicconofriNonreciprocalOscillationsPolyelectrolyte2023},~\citet{boiardiMinimalActuationControl2024},~\citet{boiardiBreakingLeftrightSymmetry2024}, and~\citet{boiardi2026three} to describe the active response of PEH filaments to an external electric field and is supplemented with the initial condition $\kappa (\cdot, 0) = \kappa_0$.
While the left-hand side of~\eqref{eq:evLaw} represents a standard relaxation dynamics with time scale $\tau$, the forcing term on the right-hand side takes into account the actual configuration of the shell through the local orientation of its unit normal relative to the applied electric field.
Specifically, $\kappa$ vanishes when the shell normal is orthogonal to the electric field, and is driven towards $\mp\kappa_E$ depending on whether $\mathbf{a}_3$ points along or against $\mathbf{e}$, with intermediate orientations leading to values in between.
This mechanical response of PEH shells refers to the polyacrylamide-co-sodium acrylate hydrogels employed for the experiments reported in~\citet{cicconofriNonreciprocalOscillationsPolyelectrolyte2023}, \citet{boiardiMinimalActuationControl2024}, and in Section~\ref{sec:experiments}.
In particular, the sign convention in~\eqref{eq:evLaw} is chosen so that the shell bends towards the direction of the applied electric field and reflects the anionic nature of the specific hydrogel considered, see Fig.~\ref{fig:GelActivity}.

\subsection{Viscous interactions}
\label{sec:nonlinearModel:viscous}

We consider the PEH shell to be neutrally buoyant and freely suspended in an electrolytic solution. We assume low-Reynolds-number hydrodynamics, so that the surrounding fluid flow is governed by the Stokes equations~\citep{happelLowReynoldsNumber1991}.
In this regime, simplified hydrodynamic theories have been developed to effectively capture viscous interactions involving slender filaments~\citep{lauga2009hydrodynamics} and ribbons~\citep{koens2016slender}, with applications to microscale biological systems.
To the best of our knowledge, however, no reduced theory is currently available to describe the viscous interaction of deformable shells in the Stokes regime.

Inspired by local resistive force theories derived for slender bodies~\citep{lauga2009hydrodynamics}, we approximate the viscous interactions between the shell and the surrounding fluid through a local drag law. In particular,
the viscous drag force per unit current surface is assumed to take the form
\begin{equation}
	\label{eq:viscDrag}
	\mathbf{d}
	= - \left[ \mu_1 \, \frac{\mathbf{a}_1 \otimes \mathbf{a}_1}{||\mathbf{a}_1||^2} + \mu_2 \, \frac{\mathbf{a}_2 \otimes \mathbf{a}_2}{||\mathbf{a}_2||^2} + \mu_3 \, \mathbf{a}_3 \otimes \mathbf{a}_3 \right] \partial_t \mathbf{u} ,
\end{equation}
where $\mu_i$ are resistive coefficients accounting for the local drag exerted by the fluid on the shell when its velocity has components along the directions defined by $\mathbf{a}_i$.
Notice that, since $|| \mathbf{a}_3 || = 1$, no normalization is required for the normal component.
The coefficients $\mu_i$ are expected to depend on the fluid dynamic viscosity $\mu$ and on geometric features of the shell, and must be specified \emph{ad hoc} for the particular shell geometry under consideration.

\subsection{Balance equations}
\label{sec:nonlinearModel:eq}

We employ the principle of virtual work to derive the balance equations in variational form.
Accordingly, for every virtual displacement $\mathbf{v}$, we require the internal virtual work to equal the external virtual work.
Since the shell is free to move in a viscous fluid, the only external contribution arises from the viscous drag~\eqref{eq:viscDrag}. The displacement is sought in a space $\mathcal{W}$ of admissible fields $\mathbf{v} = v_i \mathring{\mathbf{a}}^i$, in which kinematic boundary conditions may be in general encoded.
We neglect inertia in view of the Stokes regime assumption.

The internal virtual work coincides with the first variation of the elastic energy.
Therefore, for $t \geq 0$ we seek $\mathbf{u} (\cdot \, , t) \in \mathcal{W}$ such that
\begin{equation}
	\label{eq:PVW}
	\delta U[\mathbf{u};\mathbf{v}] := \left[ \frac{\mathrm{d}}{\mathrm{d}\varepsilon} {U}(\mathbf{u} + \varepsilon \, \mathbf{v}) \right]_{\varepsilon=0} =
	\int_\Omega \mathbf{d} \cdot \mathbf{v}  \sqrt{a} \, \mathrm{d}x^1 \mathrm{d}x^2 ,
\end{equation}
for every $\mathbf{v} \in \mathcal{W}$, where the elastic energy and the drag force are defined in~\eqref{eq:shellElasticEnergy} and~\eqref{eq:viscDrag}, respectively.
The factor $\sqrt{a}$ in the RHS accounts for the fact that $\mathbf{d}$ is a force per unit current area, whereas integration is performed over the parametric domain $\Omega$.
Making the first variation of the energy explicit, the weak form~\eqref{eq:PVW} can be rewritten as
\begin{equation}
	\label{eq:PVWExp}
	\int_\Omega \left\{ N^{\alpha \beta} \, D{G}_{\alpha \beta}^k \,v_k 
	+ M^{\alpha \beta} \, D{R}_{\alpha \beta}^k \,v_k  \right\} \!\sqrt{\mathring{a}}  \, \mathrm{d}x^1 \mathrm{d}x^2
	= 
	\int_\Omega	d^i v_i \sqrt{a} \, \mathrm{d}x^1 \mathrm{d}x^2 ,
\end{equation}
for every $\mathbf{v} = v_i\mathring{\mathbf{a}}^i \in \mathcal{W}$, where $D{G}$ and $D{R}$ are the \emph{Fr\'echet derivatives} of the change of metric and change of curvature tensors, respectively.
They are third order tensors that, when applied to a vector $\mathbf{v} = v_i\mathring{\mathbf{a}}^i$, are defined in components as
\begin{equation}
	D{G}_{\alpha \beta}^k \, v_k = \delta G_{\alpha \beta}[\mathbf{u};\mathbf{v}] , \qquad
	D{R}_{\alpha \beta}^k \, v_k =\delta R_{\alpha \beta}[\mathbf{u};\mathbf{v}] .
\end{equation}
The variational formulation~\eqref{eq:PVWExp} of the balance equations is supplemented with the initial condition $\mathbf{u} (\cdot, 0) = \mathbf{u}_0$.

We do not derive the corresponding strong form of~\eqref{eq:PVWExp}, since the subsequent stability analysis and nonlinear numerical simulations are performed using the finite element method, which relies on the weak form of the governing equations.
The interested reader may find strong forms of similar formulations in the relevant literature, for example in~\citet{efratiElasticTheoryUnconstrained2009} or in~\citet{yinBiochemomechanicalTheoryActive2021}. 

\subsection{Nondimensionalization}
\label{sec:nonlinearModel:nonDim}

To identify the relevant dimensionless groups, we recast the governing equations in dimensionless form.
Let $L_1$ and $L_2$ be two characteristic lengths of the parametric domain, $\Omega = \Omega (L_1,L_2)$.
We scale lengths by $L_1$, curvatures by $L_1^{-1}$, and time by $\tau$, the relaxation time of the spontaneous curvature introduced in~\eqref{eq:evLaw}.
Accordingly, we denote dimensionless variables and unknowns fields with a superimposed tilde, such that
\begin{equation}
	x^\alpha = L_1 \tilde{x}^\alpha , \qquad
	t = \tau \tilde{t} , \qquad
	u_i = L_1 \tilde{u}_i , \qquad
    v_i = L_1 \tilde{v}_i , \qquad
	\kappa = L_1^{-1} \tilde{\kappa} .
\end{equation}
We assume that the components of the map $\mathring{f}$ have dimensions of length. With this choice, the basis vectors $\mathring{\mathbf{a}}_i$ and $\mathbf{a}_i$ defined in Section~\ref{sec:nonlinearModel:geometry} are  dimensionless, as is the change of metric tensor $G_{\alpha \beta}$, whereas the dimensionless change of curvature tensor is $\tilde{R}_{\alpha \beta} = L_1 R_{\alpha \beta}$. Then,
the elastic energy~\eqref{eq:shellElasticEnergy} can be written as $U = Y L_1^2 h \, \tilde{U}$, where
\begin{equation}
\label{eq:energyNonDim}
	\tilde{U}
	= \frac{1}{2} \int_{\tilde{\Omega}} \left( \tilde{N}^{\alpha \beta}  G_{\alpha \beta} + \tilde{M}^{\alpha \beta} \tilde{R}_{\alpha \beta}  \right) \sqrt{\mathring{a}} \, \mathrm{d}\tilde{x}^1 \mathrm{d}\tilde{x}^2 
\end{equation}
is its dimensionless counterpart, and $\tilde{\Omega} = \tilde{\Omega} (L_2/L_1)$ is the dimensionless parametric domain in coordinates $\tilde{x}^\alpha$.
Here
\begin{gather}
\label{eq:NonLinForcesMomentsNonDim}
	\tilde{N}^{\alpha \beta} = \tilde{\mathcal{A}}^{\alpha \beta \gamma \delta} G_{\gamma \delta} \,, \quad
	\tilde{M}^{\alpha \beta} = \frac{1}{12}\left(\frac{h}{L_1}\right)^2 \tilde{\mathcal{A}}^{\alpha \beta \gamma \delta} \tilde{R}_{\gamma \delta} ,
\end{gather}
are the dimensionless internal membrane  forces and moments, with
\begin{equation}
	\tilde{\mathcal{A}}^{\alpha \beta \gamma \delta} = \frac{1}{1+\nu} \left( \frac{\nu}{1-\nu} \mathring{a}^{\alpha \beta} \mathring{a}^{\gamma \delta} + \mathring{a}^{\alpha \gamma} \mathring{a}^{\beta \delta} \right)  ,
\end{equation}
the dimensionless elasticity tensor.
It is then natural to introduce the two dimensionless groups $\lambda = L_2/L_1$ and $\eta = h/L_1$ that encode the geometry of the system.

As for the resistive coefficients $\mu_i$ introduced in~\eqref{eq:viscDrag}, they depend on the dynamic viscosity $\mu$ of the surrounding fluid and on the geometric parameters of the shell, \ie, $\mu_i = \mu_i (\mu, L_1, L_2, h)$.
By Buckingham $\Pi$ theorem~\citep{buckingham1914physically}, they can be written as
\begin{equation} \label{eq:pi_therom_visc}
	\mu_i = \frac{\mu}{L_2} \zeta_i (\lambda, \eta) ,
\end{equation}
where  $\zeta_i =  \zeta_i (\lambda, \eta)$ are dimensionless functions of $\lambda$ and $\eta$.
The choice of dividing by $L_2$ becomes apparent from the explicit form of the coefficients detailed in~\ref{sec:appViscPlate} and~\ref{sec:appViscCyl}.
The viscous drag is then $\mathbf{d} = [\mu L_1 / ( L_2 \tau) ] \, \tilde{\mathbf{d}}$, with
\begin{equation}
\label{eq:viscDragNonDim}
	\tilde{\mathbf{d}}
	= - \left[ \zeta_1 \, \frac{\mathbf{a}_1 \otimes \mathbf{a}_1}{||\mathbf{a}_1||^2} + \zeta_2 \, \frac{\mathbf{a}_2 \otimes \mathbf{a}_2}{||\mathbf{a}_2||^2} + \zeta_3 \, \mathbf{a}_3 \otimes \mathbf{a}_3 \right] \partial_{\tilde{t}} \tilde{\mathbf{u}} ,
\end{equation}
so that the external virtual work reads
\begin{equation}
	\int_\Omega \mathbf{d} \cdot \mathbf{v}  \sqrt{a} \, \mathrm{d}x^1 \mathrm{d}x^2 
	= \frac{\mu L_1^4}{L_2 \tau} \int_{\tilde{\Omega}} \tilde{\mathbf{d}} \cdot \tilde{\mathbf{v}} \sqrt{a} \, \mathrm{d}\tilde{x}^1 \mathrm{d}\tilde{x}^2 .
\end{equation}

The ratio between the coefficients multiplying the dimensionless elastic energy, and hence its variation, and the dimensionless external virtual work is
\begin{equation}
	\frac{Y L_1^2 h}{\mu L_1^4 / ( L_2 \tau)} = \frac{(L_2/L_1) (h/L_1)}{\mu/ (Y \tau)} =: \frac{\lambda \, \eta}{\xi} ,
\end{equation}
which allows us to rewrite the weak formulation~\eqref{eq:PVWExp} in dimensionless form as
\begin{equation}
\label{eq:PVWExpNonDim}
	\frac{\lambda \, \eta}{\xi} \int_{\tilde{\Omega}} \left\{ \tilde{N}^{\alpha \beta}  DG_{\alpha \beta}^k \,  \tilde{v}_k
	+ \tilde{M}^{\alpha \beta}  D\tilde{R}_{\alpha \beta}^k \,\tilde{v}_k  \right\} \sqrt{\mathring{a}} \, \mathrm{d}\tilde{x}^1 \mathrm{d}\tilde{x}^2
	= \int_{\tilde{\Omega}} \tilde{d}^i  \tilde{v}_i  \sqrt{a} \, \mathrm{d}\tilde{x}^1 \mathrm{d}\tilde{x}^2 ,
\end{equation}
for every $\tilde{\mathbf{v}} = \tilde{v}_i \mathring{\mathbf{a}}^i \in \tilde{\mathcal{W}}$. Finally, the dimensionless counterpart of the evolution equation~\eqref{eq:evLaw} for the spontaneous curvature reads
\begin{equation}
\label{eq:evLawNonDim}
	\partial_{\tilde{t}}\, \tilde{\kappa} + \tilde{\kappa} = - \chi \, \mathbf{e} \cdot \mathbf{a}_3 ,
\end{equation}
where $\chi = L_1 \kappa_E$.
We collect the relevant dimensionless groups in Table~\ref{tab:adm_par}, and recall that $\tilde{\Omega} = \tilde{\Omega}(\lambda)$ and, in general, $\zeta_i = \zeta_i (\lambda, \eta)$.
With a slight abuse of notation, hereafter we denote dimensionless quantities by the same symbols as their dimensional counterparts and omit the superimposed tildes.
\begin{table}[t]
	\centering
	\footnotesize
	\begin{tabular}{ccl}
		\toprule
		~~~~~~Model parameter~~~~~~ & ~~~~~~Parameter expression~~~~~~ & ~~~~~~Parameter description~~~~~~\\
		\midrule
		$\nu$       &       -    &   material Poisson's ratio \\[1mm]
		$\lambda$   &       $L_2/L_1$    &   aspect ratio of the parametric domain \\[1mm]
		$\eta$      &       $h/L_1$    &   thickness over characteristic length \\[1mm]
		$\xi$       &       $\mu/(Y \tau)$    &   passive/active time ratio \\[1mm]
		$\chi$      &       $L_1 \kappa_E$    &   external driving stimulus \\
		\bottomrule
	\end{tabular}
	\medskip
	\caption{Independent dimensionless model parameters as introduced by recasting the governing equations in dimensionless form.}
	\label{tab:adm_par}
\end{table}
%

%%%%%%%%%%%%%%%%%%%%%%%%%%%%%%%%%%%%%%%%%%%%%%%%%%%%%%%%%%%%%%%%

\section{Elliptic active plates}
\label{sec:plates}

\begin{figure}[t]
    \centering
    \includegraphics[width=1\linewidth]{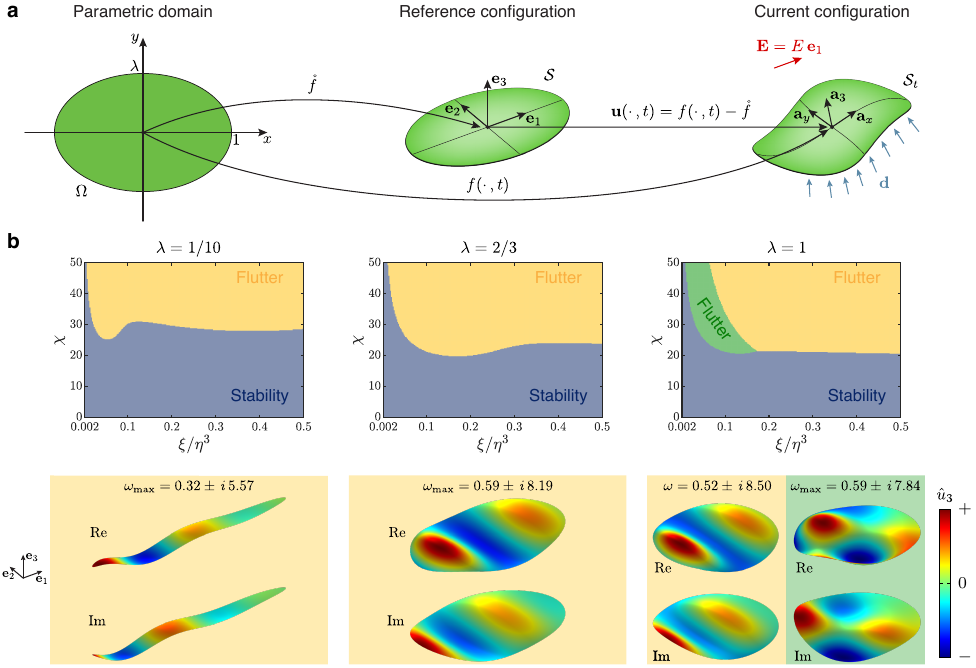}
    \caption{
    (a) Parametric domain $\Omega$, reference configuration $\mathcal{S}$, and current configuration $\mathcal{S}_t$ of an elliptic active plate.
    (b) Results of the linear stability analysis.
    The first row shows the stability regions of the system while varying the parameters $\xi/\eta^3 \in [0.002, 0.5]$, $\chi \in [0,50]$, and $\lambda \in \{1/10, 2/3, 1\}$.
    Blue regions correspond to parameters for which the system is stable, while yellow and green regions indicate parameters for which the system is in the flutter regime.
    The second row reports the leading fluttering eigenvalues and depicts the real and imaginary parts of the corresponding eigenmodes $\hat{u}_3$ for $(\xi/\eta^3, \chi) = (0.054, 36.768)$, varying $\lambda \in \{1/10, 2/3, 1\}$.
    For the circular plate, $\lambda = 1$, the colors distinguish between flutter associated with \emph{mode 1} (yellow) and \emph{mode 2} (green) in both rows.
    }
    \label{fig:PlateGeometry}
\end{figure}

In this section, we specialize the model to active elliptic plates subjected to an external field aligned with their major axis.
We identify the coordinates as $(x^1, x^2) \equiv (x, y)$ and adopt the convention that Greek indices range over $\{x, y\}$ throughout this section.
The mid-surface of the reference elliptic plate is identified by
\begin{equation}
    \mathring{f} (x,y) = (x,y,0)\,, \qquad (x,y) \in \Omega \,,
\end{equation}
where the coordinates refer to the orthonormal basis $\{ \mathbf{e}_1, \mathbf{e}_2, \mathbf{e}_3\}$.
The parametric domain $\Omega$ corresponds to an ellipse with semi-axes $1$ and $\lambda \leq 1$, as depicted in Fig.~\ref{fig:PlateGeometry}a.
The basis vectors are $\mathring{\mathbf{a}}_x = \mathbf{e}_1$ , $\mathring{\mathbf{a}}_y = \mathbf{e}_2$ and $\mathring{\mathbf{a}}_3 = \mathbf{e}_3$ ,
so that the direction of the electric field is $\mathbf{e} = \mathbf{e}_1$.
The covariant components of the reference fundamental forms are
\begin{equation}
    [\mathring{a}_{\alpha \beta}] = 
    \begin{bmatrix}
        1 & 0 \\
        0 & 1
    \end{bmatrix}, 
    \qquad
    [\mathring{b}_{\alpha \beta}] = 
    \begin{bmatrix}
        0 & 0 \\
        0 & 0
    \end{bmatrix},
\end{equation}
signifying that the reference metric is Euclidean and the undeformed configuration is intrinsically flat.
Since $\mathbf{\mathring{a}}$ coincides with the identity tensor, the distinction between covariant and contravariant components vanishes, and we retain the index notation solely for consistency with the previous section.
From~\eqref{eq:refTensors}, the natural metric and curvature tensors in components read 
\begin{equation}
    [\bar{a}_{\alpha \beta}] = 
    \begin{bmatrix}
        1 & 0 \\
        0 & 1
    \end{bmatrix}, 
    \qquad
    [\bar{b}_{\alpha \beta}] = 
    \begin{bmatrix}
        - \kappa & 0 \\
        0 & - \kappa
    \end{bmatrix} .
\end{equation}
These reveal the incompatibility of the natural, stress-free configuration: the natural metric is Euclidean, and therefore imposes a vanishing Gaussian curvature, whereas $\mathrm{det}(\bar{b}^\alpha_\beta) = \kappa^2$. Hence, the Gauss equation cannot be satisfied unless $\kappa = 0$, so that no surface exists with $\bar{\mathbf{a}}$ and $\bar{\mathbf{b}}$ as fundamental forms.
The Euclidean reference metric implies that the components of forces and moments in~\eqref{eq:NonLinForcesMomentsNonDim} for a plate read
\begin{gather}
    N^{xx} =  \frac{1}{1-\nu^2} \left( G_{xx} + \nu \, G_{yy} \right) , 
    \quad N^{yy} =  \frac{1}{1-\nu^2} \left( G_{yy} + \nu \, G_{xx} \right) ,
    \quad N^{xy} =  N^{yx} = \frac{1}{1+\nu} G_{xy} , \\
    M^{xx}  =  \frac{\eta^2}{12(1-\nu^2)} \left( R_{xx} + \nu \, R_{yy} \right) ,
    \quad M^{yy} =  \frac{\eta^2}{12(1-\nu^2)} \left( R_{yy} + \nu \, R_{xx} \right) , 
    \quad M^{xy} = M^{yx} = \frac{\eta^2}{12(1+\nu)} R_{xy} .
\end{gather}
The viscous coefficients $\zeta_i$ are derived from the resistive coefficients of an ellipsoid with semi-axes $\{1,\lambda,\eta/2\}$ in the limit of $\eta \to 0^+$, as detailed in~\ref{sec:appViscPlate}.
Consequently, within this section, the coefficients depend solely on the ellipse aspect ratio, $\zeta_i = \zeta_i (\lambda)$.

\subsection{Linear stability analysis}
\label{sec:plates:stability}

To study the stability of the system in response to the external electric field, we linearize the governing equations around the trivial equilibrium characterized by $\mathbf{u}_\mathrm{eq} = \mathbf{0}$ and $\kappa_\mathrm{eq} = 0$.
We consider perturbations of this equilibrium and set $\mathbf{u} = \varepsilon \, \mathbf{u}^l + o(\varepsilon)$ and
$\kappa = \varepsilon \, \kappa^l  + o(\varepsilon)$, where $0< \varepsilon \ll 1$ is a small parameter.
Retaining first order terms in $\varepsilon$, the linearized counterpart of the weak balance equations~\eqref{eq:PVWExpNonDim} are~\citep{IntroductionDifferentialGeometry2005}
\begin{equation}
\label{eq:PVWExpNonDimLin}
   \frac{\lambda \, \eta}{\xi} \int_\Omega \left\{ n^{\alpha \beta} (\mathbf{u}^l) \,\gamma_{\alpha \beta}(\mathbf{v})
    + m^{\alpha \beta} (\mathbf{u}^l) \,\rho_{\alpha \beta}(\mathbf{v}) \right\}  \, \mathrm{d}x \mathrm{d}y 
    = \int_\Omega d^i_l v_i  \, \mathrm{d}x \mathrm{d}y ,
\end{equation}
where
\begin{equation}
\label{eq:plateLinStrains}
    \gamma_{\alpha \beta} (\mathbf{v}) =  
    \begin{bmatrix}
        \partial_x v_x & \left( \partial_x v_y + \partial_y v_x \right) / 2 \\[1mm]
        \left( \partial_x v_y + \partial_y v_x \right) / 2 & \partial_y v_y
    \end{bmatrix}, 
    \qquad
    \rho_{\alpha \beta}(\mathbf{v}) =
    \begin{bmatrix}
        \partial_x^2 v_3 & \partial_{xy} v_3 \\[1mm]
        \partial_{xy} v_3 & \partial_y^2 v_3
    \end{bmatrix}
\end{equation}
are the linearized change of metric and change of curvature tensors for a virtual configuration identified by a virtual displacement $\mathbf{v} = v_i \mathring{\mathbf{a}}^i$ with respect to the undeformed one, and 
\begin{alignat}{2}
    n^{xx} (\mathbf{u}^l) & =  \frac{1}{1-\nu^2} \left( \gamma_{xx} (\mathbf{u}^l) + \nu \, \gamma_{yy} (\mathbf{u}^l)\right) , 
    \qquad & m^{xx} (\mathbf{u}^l)  &=  \frac{\eta^2}{12(1-\nu^2)} \left[ \left(\rho_{xx} (\mathbf{u}^l) + \kappa^l \right) + \nu \left( \rho_{yy} (\mathbf{u}^l) + \kappa^l \right) \right] , 
    \label{eq:plateLinStress1} \\
    n^{yy} (\mathbf{u}^l) & =  \frac{1}{1-\nu^2} \left( \gamma_{yy} (\mathbf{u}^l) + \nu \, \gamma_{xx} (\mathbf{u}^l)\right) ,
    \qquad & m^{yy} (\mathbf{u}^l) &=  \frac{\eta^2}{12(1-\nu^2)} \left[ \left(\rho_{yy} (\mathbf{u}^l) + \kappa^l \right) + \nu \left( \rho_{xx} (\mathbf{u}^l) + \kappa^l \right) \right] , 
    \label{eq:plateLinStress2} \\
    n^{xy} (\mathbf{u}^l )& = \frac{1}{1+\nu} \gamma_{xy} (\mathbf{u}^l),  
    \qquad & m^{xy} (\mathbf{u}^l) &= \frac{\eta^2}{12(1+\nu)} \rho_{xy} (\mathbf{u}^l) ,
    \label{eq:plateLinStress3}
\end{alignat}
are the linearized internal forces and moments corresponding to the displacement $\mathbf{u}^l$ with respect to stress-free configuration. 
We highlight that $\kappa^l$ is not present in~\eqref{eq:plateLinStrains} while it is in~\eqref{eq:plateLinStress1}--\eqref{eq:plateLinStress2}, since virtual variations are performed in the displacement only.
The linearized viscous force is
\begin{equation}
   \mathbf{d}_l = 
   d^i_l \, \mathring{\mathbf{a}}_i = 
   - \zeta_x \, \partial_t u_x^l \, \mathring{\mathbf{a}}_x - \zeta_y \,\partial_t u_y^l \, \mathring{\mathbf{a}}_y - \zeta_3 \, \partial_t u_3^l \, \mathring{\mathbf{a}}_3 ,
\end{equation}
and the linearized counterpart of the evolution equation~\eqref{eq:evLawNonDim} reads
\begin{equation}
\label{eq:evLawNonDimLin}
   \partial_t \kappa^l + \kappa^l = \chi \,\partial_x u_3^l .
\end{equation}

The weak form~\eqref{eq:PVWExpNonDimLin} is tested with $\mathbf{v} = v_x \, \mathring{\mathbf{a}}_x$, $\mathbf{v} = v_y \, \mathring{\mathbf{a}}_y$, and $\mathbf{v} = v_3 \, \mathring{\mathbf{a}}_3$, respectively, to obtain the three equations
\begin{align}
   \int_\Omega \bigg\{ n^{xx} (\mathbf{u}^l) \, \partial_x v_x + n^{xy} (\mathbf{u}^l) \,\partial_y v_x + \frac{\zeta_x \xi}{\lambda \eta} \, \partial_t u_x^l \, v_x \bigg\}  \, \mathrm{d}x \mathrm{d}y &= 0 ,
   \label{eq:BalanceXPlateLin} \\[2mm]
   \int_\Omega \bigg\{ n^{yy} (\mathbf{u}^l) \, \partial_y v_y + n^{xy} (\mathbf{u}^l) \,  \partial_x v_y + \frac{\zeta_y \xi}{\lambda \eta} \, \partial_t u_y^l \, v_y \bigg\}  \, \mathrm{d}x \mathrm{d}y &= 0 , 
   \label{eq:BalanceYPlateLin} \\[2mm]
   \int_\Omega \bigg\{ m^{xx} (\mathbf{u}^l) \, \partial_x^2 v_3 + m^{yy} (\mathbf{u}^l) \, \partial_y^2 v_3 + 2 m^{xy} (\mathbf{u}^l) \, \partial_{xy} v_3 + \frac{\zeta_3 \xi}{\lambda \eta} \, \partial_t u_3^l \, v_3 \bigg\}  \, \mathrm{d}x \mathrm{d}y &= 0 , 
   \label{eq:Balance3PlateLin}
\end{align}
holding for any suitable test functions $v_x, \, v_y, \, v_3$. Similarly, the evolution law~\eqref{eq:evLawNonDimLin} is recast in weak form as
\begin{equation}
    \label{eq:EVLAW_WEAK}
    \int_\Omega \bigg\{\left[\partial_t \kappa^l + \kappa^l - \chi \, \partial_x u_3^l\right]\psi \bigg\}  \, \mathrm{d}x \mathrm{d}y = 0 ,
\end{equation}
where $\psi$ is the test function for the active curvature.

As is known from classical plate theory, membrane and flexural problems decouple in the linear regime. Indeed, the membrane forces and the bending moments depend only on the in-plane displacements $u_\alpha^l$ and on the transverse displacement $u_3^l$, respectively~\citep{timosenkoTheoryPlatesShells1996}.
Since no kinematic constraints are imposed on the plate, its activity is coupled to the bending moments only, and the external loads are of viscous nature, the membrane problem~\eqref{eq:BalanceXPlateLin}-\eqref{eq:BalanceYPlateLin} is solved trivially by vanishing in-plane displacements.
Hence, the stability analysis reduces to solve the flexural equation~\eqref{eq:Balance3PlateLin} and the evolution equation~\eqref{eq:EVLAW_WEAK} in the unknowns $u_3^l$ and $\kappa^l$.
In particular, it requires to seek time-harmonic solutions of the form
\begin{equation}
\label{eq:TH_PLATE}
    u_3^l (x,y,t) = \mathrm{exp} (\omega \, t) \, \hat{u}_3 (x,y) , \qquad
    \kappa^l (x,y,t) = \mathrm{exp} (\omega \, t) \, \hat{\kappa} (x,y) ,
\end{equation}
and to study the ensuing eigenvalue problem to characterize the angular frequency $\omega$ and the corresponding eigenfunctions $\hat{u}_3$ and $\hat{\kappa}$ in relation to the model parameters.
We do so numerically, as detailed in~\ref{sec:appDetailsPlatesStab}, and report in the first row of  Fig.~\ref{fig:PlateGeometry}b the computed stability maps for fixed values of the ellipse aspect ratio $\lambda \in \{ 1/10, \, 2/3, \, 1 \}$ and for $\nu = 0.3$, see~\ref{sec:appElasticPEH} for details on the parameter choice.
Then, by recalling that $\zeta_3 = \zeta_3 (\lambda)$, the stability of the system is completely determined by $\chi$ and the ratio $\xi / \eta^3$. For all $\lambda$, we explore the parameter ranges $\chi \in [0, 50]$ and $\xi / \eta^3 \in [0.002, 0.5]$.
We recall that a change in the sign of $\chi$ corresponds to a reversal of the electric field, leading to solutions that are reflected along the ellipse major axis.

We denote by $\omega_{\mathrm{max}} = \alpha \pm \mathrm{i} \, \beta$ the eigenvalue with the largest real part. Then, in the maps of Fig.~\ref{fig:PlateGeometry}b, blue identifies \emph{stability} regions $(\alpha \leq 0)$, whereas
yellow and green identify \emph{flutter} regions  $(\alpha > 0,\,\, \beta \neq 0)$, so that a perturbation of the equilibrium leads to oscillations of increasing amplitude. We notice that the critical value of $\chi$ at the onset of flutter depends on $\xi / \eta^3$, with a stronger dependence at smaller values of this parameter. 
We report in the second row of Fig.~\ref{fig:PlateGeometry}b the real and imaginary parts of the oscillatory eigenmodes $\hat{u}_3$ and the corresponding eigenvalues $\omega$ for $\chi = 36.768$ and for $\xi / \eta^3 = 0.054$. We anticipate that these parameters  are relevant to the nonlinear simulations and the experiments later discussed, see~\ref{sec:appElasticPEH}.
Returning to the stability maps, flutter regions are color-coded according to the leading oscillatory mode. For $\lambda = 1/10$ and $\lambda = 2/3$, flutter instability is associated with \emph{mode 1} (yellow shading). This exhibits curvature predominantly along the major axis, although a non-vanishing curvature is present also along the minor axis. For the case of $\lambda = 1$, we identify a parameter region where the leading eigenvalue is associated with \emph{mode 2} (green shading), characterized by a significant double curvature throughout the plate.
In this regard, we recall that membrane and flexural problems are decoupled in the linear regime, but they are not at large displacements. Consequently, we expect modes characterized by significant double curvature to be suppressed in the nonlinear setting.
We emphasize that the real and imaginary parts of the oscillatory eigenmodes are not proportional, such that their modulation in time results in nonreciprocal periodic shape changes of increasing amplitude.
Finally, we note that, for the considered parameter  range, our results also reveal \emph{divergence modes}, characterized by positive real eigenvalues $(\alpha > 0, \,\, \beta =0)$. However, these were not included in the stability maps, as nonlinear simulations reveal that they are not maintained at large displacements, with the system transitioning  towards oscillatory modes.

\subsection{Nonlinear numerical results}
\label{sec:plates:simulations}

\begin{figure}[p]
	\centering
	\includegraphics[width=\textwidth]{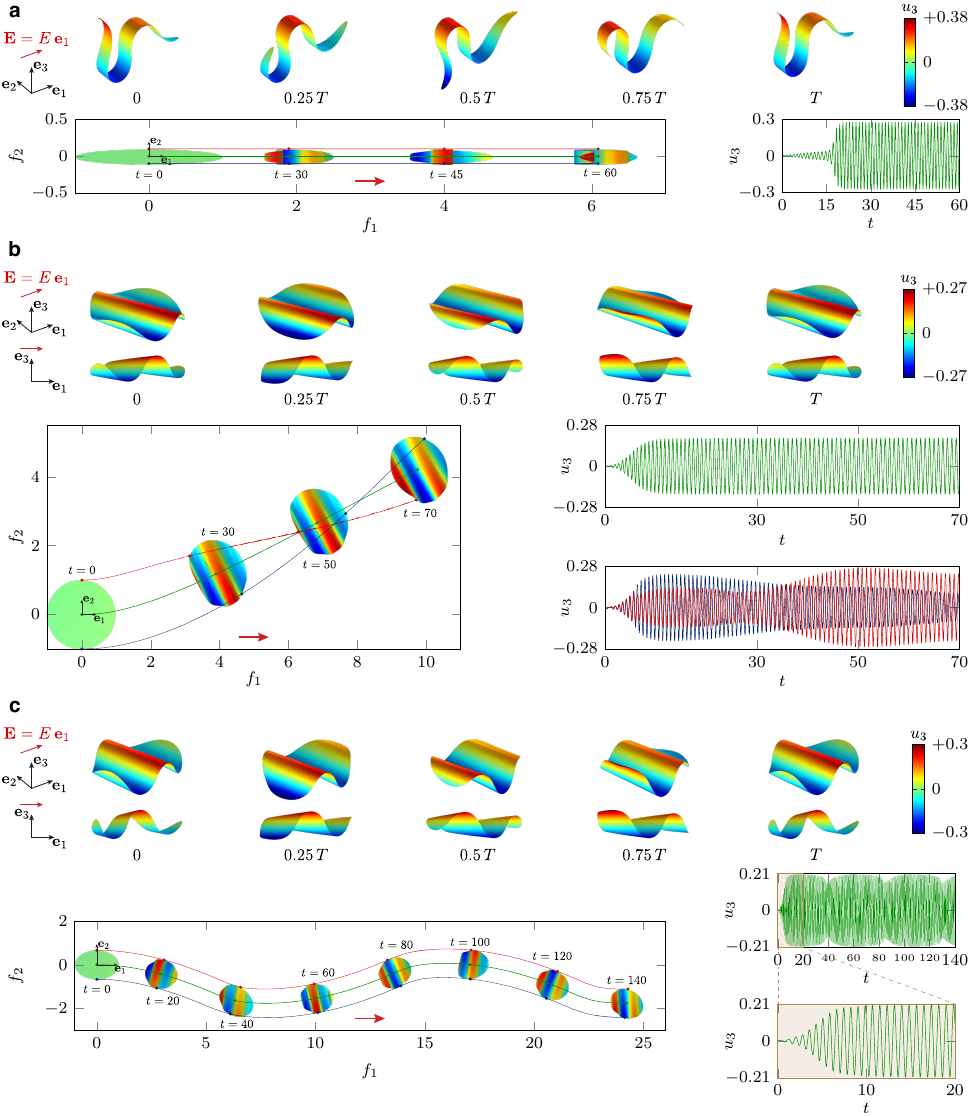}
	\caption{
		Results of nonlinear simulations with $\nu = 0.3$, $\chi = 36.768$, $\xi = 2.967 \times 10^{-8}$, $\eta = 8.171 \times 10^{-3}$, for elliptic plates with aspect ratio $\lambda = 1/10$ (a), $\lambda = 1$ (b), and $\lambda = 2/3$ (c).
		For each value of $\lambda$, the top row displays five snapshots of the deformed configuration covering one period $T$, once the limit cycle is reached.
		The bottom row shows the trajectory of the plate (left) and the time evolution of the transverse displacement for selected material points (right).
		Here, $f_1 = (f - o) \cdot \mathbf{e}_1$ and $f_2 = (f - o) \cdot \mathbf{e}_2$ are the Cartesian components of the current configuration, with $o = (0,0,0) \in \mathcal{E}$.
		All curves are color-coded and correspond to material points located at $(0,0)$ (green), $(0,\lambda)$ (red), and $(0,-\lambda)$ (blue) in the parametric domain $\Omega$.
        See also Supplementary Videos~1--3.
	} \label{fig:PlatesNonlinearResults}
\end{figure}

In this section, we investigate the dynamic behavior of active elliptic plates through nonlinear simulations, providing details on the numerical implementation of the governing equations in~\ref{sec:appDetailsPlatesNonlin}.
We report in Fig.~\ref{fig:PlatesNonlinearResults} the results of simulations performed with $\nu = 0.3$, $\chi = 36.768$, $\xi = 2.967 \times 10^{-8}$, and $\eta = 8.171 \times 10^{-3}$, see~\ref{sec:appElasticPEH}.
These parameters correspond to the eigenvalues and eigenmodes reported in the last row of Fig.~\ref{fig:PlateGeometry}b, for which the system is in the flutter regime.

We first discuss the results for a ribbon-like plate with $\lambda = 1/10$, see Fig.~\ref{fig:PlatesNonlinearResults}a. The system departs from the trivial equilibrium configuration through oscillations of increasing amplitude, which closely resemble the flutter mode reported in Fig.~\ref{fig:PlateGeometry}b, bottom-left panel. A limit cycle characterized by large amplitude  shape changes is attained at $t \approx 22.5$. By this motion of period $T \approx 1.11$, the elongated plate undergoes a net displacement in the direction of the external electric field, see Supplementary Video~1. In particular, the top row of Fig.~\ref{fig:PlatesNonlinearResults}a shows five shapes within a period, while the bottom row shows the straight trajectory of the system. The transverse displacement of the plate center highlights the oscillations of increasing amplitude, until the limit cycle is reached.
Remarkably, the imaginary part of the leading eigenvalue $\omega_\mathrm{max} = 0.32 \pm \mathrm{i} \, 5.57$ predicts a period of $2 \pi / 5.57 \approx 1.13$.
We also highlight that the presented results quantitatively match those of~\citet{boiardiMinimalActuationControl2024} for a rod-like swimmer, showing that a one-dimensional model is suitable to study the response to electric fields of PEH-based ribbon-like structures.

We next consider the limiting case of a circular plate with $\lambda = 1$. The relevant results are reported in Fig.~\ref{fig:PlatesNonlinearResults}b, see also Supplementary Video~2. In analogy to the previous case, an initial perturbation leads to oscillatory shape changes of increasing amplitude. At small amplitudes, these resemble the leading eigenmode (\emph{mode 2}) reported in Fig.~\ref{fig:PlateGeometry}b, as predicted by the linear stability analysis. However, as the oscillations increase in amplitude, this eigenmode is not sustained, as due to the competition between bending and stretching energies in the small thickness regime, and the other flutter mode (\emph{mode 1}) depicted in Fig.~\ref{fig:PlateGeometry}b prevails. 
Nevertheless, this mode is also characterized by a significant double curvature, so that the fully developed periodic motion, reached at $t \approx 10$ with period $T \approx 0.84$, exhibits shapes that cannot be directly related to those predicted by the stability analysis, unlike the ribbon-like case. The emergent dynamics
still results in an net motion, which is not in the direction of the electric field, but at a specific angle to it.
The corresponding shapes qualitatively resemble \emph{ruled surfaces}, exhibit an oscillation amplitude that continuously varies across the transverse direction, and break symmetry with respect to the direction of motion.
A closer investigation of the trajectories and transverse displacements of selected material points, highlighted in red, green, and blue in the bottom panels of Fig.~\ref{fig:PlatesNonlinearResults}b, shows that these asymmetric shape changes are accompanied by the rotation of the plate about the vertical axis. Owing to the geometric symmetry of the circular plate, this rotation cannot be inferred from the plate configurations alone.

Finally, we discuss the intermediate case of an elliptic plate with $\lambda = 2/3$, see Fig.~\ref{fig:PlatesNonlinearResults}c. The transverse displacement of the plate center is depicted in the bottom-right panel of the figure for $t \lesssim 20$, showing  oscillations that progressively increase in amplitude.
As for the previous case, also the elliptic plate exhibits asymmetric shapes changes at larger times, as reported in the top panel. This asymmetry  leads again to a rotation of the system about the vertical axis.
However, unlike the circular plate case, this rotation does not occur at a constant angular velocity. As the plate reorients, it presents a varying projected length to the electric field, so that the motion continuously adapts in time, giving rise to curved swimming paths, as highlighted by the trajectories of the material points shown in red, green, and blue, see Supplementary Video~3.
More specifically, the amplitude of the oscillations  decreases over the time interval $20 \lesssim t \lesssim 40$, bringing the system close to a stalled state. Subsequently, as the field magnitude is sufficiently above the critical value of flutter, the oscillations increase again in amplitude, the direction of rotation reverses, and the plate resumes its motion.  
This complex dynamics repeats periodically, giving rise to a long-range swing-like motion, as evidenced by the trajectories in the bottom panel and by the transverse displacement of the center in the right panel of Fig.~\ref{fig:PlatesNonlinearResults}c.
We identify two different time scales: the \emph{actual} period of the complex dynamics described above, which is of approximately 90 time units from $t \approx 40$ to $t \approx 130$, and the \emph{short-time} period $T$, which is comparable to that of the other two case studies and ranges from $T \approx 0.81$ to $T \approx 0.84$.

%%%%%%%%%%%%%%%%%%%%%%%%%%%%%%%%%%%%%%%%%%%%%%%%%%%%%%%%%%%%%%%%

\section{Cylindrical active shells}
\label{sec:cylindricalShells}

In this section, we specialize the model to  cylindrical active shells with circular cross-section, subjected to an electric field directed along their axis.
We identify the coordinates as $(x^1,x^2) \equiv (s,z)$ and adopt the convention that, throughout this section, Greek indices range over $\{s, z\}$.
\begin{figure}[th!]
    \centering
    \includegraphics[width=1\linewidth]{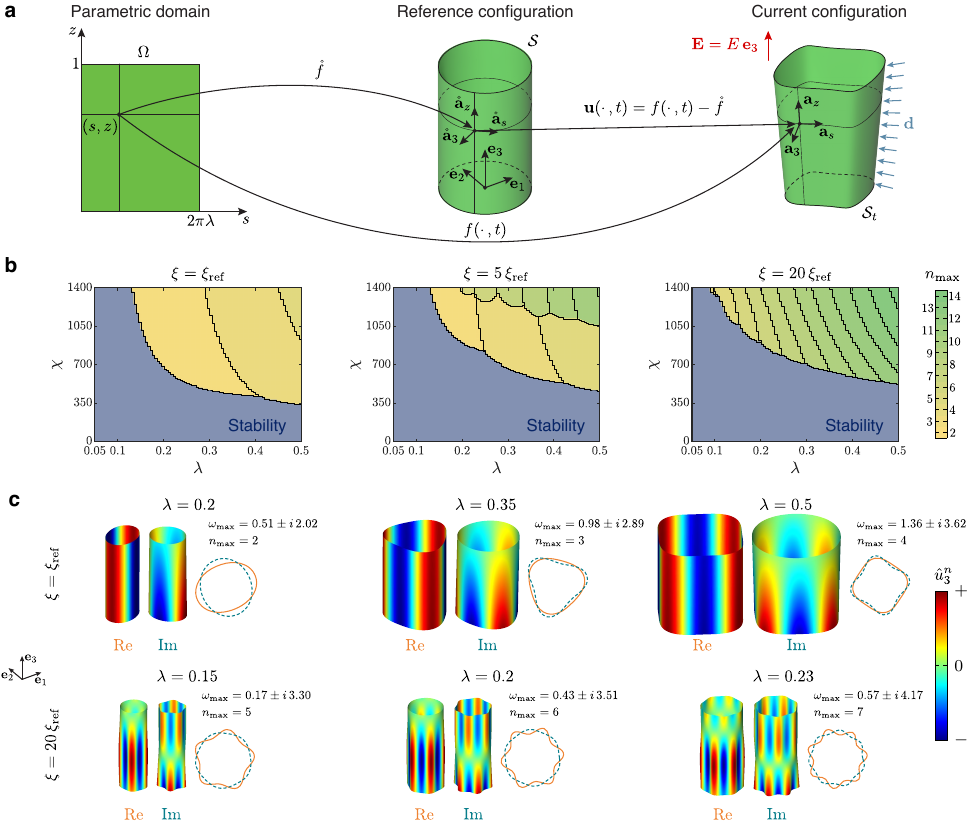}
    \caption{
    (a) Parametric domain $\Omega$, reference configuration $\mathcal{S}$, and current configuration $\mathcal{S}_t$ of a cylindrical active shell with circular cross-section.
    (b) Results of the linear analysis, showing the stability regions of the system while varying the parameters  $\lambda \in [0.05, 0.5]$, $\chi \in [0,1400]$, and $\xi = \{\xi_\mathrm{ref}, 5 \,\xi_\mathrm{ref}, 20 \, \xi_\mathrm{ref}\}$, with $\xi_\mathrm{ref} = 2.967 \times 10^{-8}$.
    Blue regions correspond to parameters for which the system is stable, while regions for which the system is in the flutter regime are color-coded from yellow to green according to the circumferential mode $n_\mathrm{max}$ corresponding to the leading eigenvalue.
    (c) Leading fluttering eigenvalues and real and imaginary part of the corresponding eigenmodes in a three-dimensional view and in a top view corresponding to $ z = 1/2$. The first row depicts results for $\xi = \xi_\mathrm{ref}$ and $\lambda = \{ 0.2, 0.35, 0.5 \}$; the second row depicts results for $\xi = 20 \, \xi_\mathrm{ref}$ and $\lambda = \{ 0.15, 0.2, 0.23 \}$.
    }
    \label{fig:CylGeometry}
\end{figure}

We consider a cylindrical shell of height $1$ and radius $\lambda$. With reference to the orthonormal basis $\{ \mathbf{e}_1, \mathbf{e}_2, \mathbf{e}_3\}$, its undeformed mid-surface is identified by
\begin{equation}
    \mathring{f} (s,z) = \left(\lambda \cos \frac{s}{\lambda},\lambda \sin \frac{s}{\lambda}, z \right) , \qquad (s, z) \in \Omega ,
\end{equation}
where the parametric domain is $\Omega = [0,2\pi \lambda] \times [0,1]$, see Fig.~\ref{fig:CylGeometry}a.
Then the basis vectors are
\begin{equation}
    \mathring{\mathbf{a}}_s = -\sin \frac{s}{\lambda} \, \mathbf{e}_1 + \cos \frac{s}{\lambda} \, \mathbf{e}_2 , \qquad
    \mathring{\mathbf{a}}_z = \mathbf{e}_3 , \qquad 
    \mathring{\mathbf{a}}_3 = \cos \frac{s}{\lambda} \mathbf{e}_1 + \sin \frac{s}{\lambda} \mathbf{e}_2 ,
\end{equation}
so that the electric field is directed along $\mathbf{e} = \mathbf{e}_3$.
The covariant components of the reference fundamental forms are
\begin{equation}
    [\mathring{a}_{\alpha \beta}] = 
    \begin{bmatrix}
        1 & 0 \\
        0 & 1
    \end{bmatrix}, 
    \qquad
    [\mathring{b}_{\alpha \beta}] = 
    \begin{bmatrix}
        - 1/\lambda & 0 \\
        0 & 0
    \end{bmatrix}.
\end{equation}
Thus, the reference metric is Euclidean and the principal curvatures of the undeformed shell are $-1/\lambda$ and $0$.
Since $\mathring{\mathbf{a}}$ is the identity tensor, no distinction between covariant and contravariant components is required.
From~\eqref{eq:refTensors}, the natural metric and curvature tensors in components read 
\begin{equation}
    [\bar{a}_{\alpha \beta}] = 
    \begin{bmatrix}
        1 & 0 \\
        0 & 1
    \end{bmatrix}, 
    \qquad
    [\bar{b}_{\alpha \beta}] = 
    \begin{bmatrix}
        - 1 / \lambda - \kappa & 0 \\
        0 & - \kappa
    \end{bmatrix} .
\end{equation}
As for the case of  plates, the Euclidean reference metric implies that the components of forces and moments in~\eqref{eq:NonLinForcesMomentsNonDim} for a cylindrical shell are given by
\begin{gather}
    N^{ss} =  \frac{1}{1-\nu^2} \left( G_{ss} + \nu \, G_{zz} \right) , 
    \quad N^{zz} =  \frac{1}{1-\nu^2} \left( G_{zz} + \nu \, G_{ss} \right) ,
    \quad N^{sz} =  N^{zs} = \frac{1}{1+\nu} G_{sz} , \\
    M^{ss}  =  \frac{\eta^2}{12(1-\nu^2)} \left( R_{ss} + \nu \, R_{zz} \right) ,
    \quad M^{zz} =  \frac{\eta^2}{12(1-\nu^2)} \left( R_{zz} + \nu \, R_{ss} \right) , 
    \quad M^{sz} = M^{zs} = \frac{\eta^2}{12(1+\nu)} R_{sz} .
\end{gather}

Regarding the  dimensionless resistive coefficients $\zeta_i$, we take them to be constant. As justified in~\ref{sec:appViscCyl}, we set $\zeta_s = \zeta_z = 32 / (3 \pi)$ and $\zeta_3 = 16 / \pi$  throughout this section.

\subsection{Linear stability analysis}
\label{sec:cylindricalShells:stability}

We now investigate the stability of the system about the trivial equilibrium $\mathbf{u}_\mathrm{eq} = \mathbf{0}$ and $\kappa_\mathrm{eq} = 0$.
To this end, we consider perturbations of the form $\mathbf{u} = \varepsilon \, \mathbf{u}^l + o(\varepsilon)$ and $\kappa = \varepsilon \, \kappa^l + o(\varepsilon)$, with $0< \varepsilon \ll 1$ a small parameter, leading to the linearization of the balance equations~\eqref{eq:PVWExpNonDim}
\begin{equation}
\label{eq:PVWExpNonDimLinCyl}
   \frac{\lambda \, \eta}{\xi} \int_\Omega \left\{ n^{\alpha \beta} (\mathbf{u}^l) \,\gamma_{\alpha \beta}(\mathbf{v})
    + m^{\alpha \beta} (\mathbf{u}^l) \,\rho_{\alpha \beta}(\mathbf{v}) \right\}  \mathrm{d}s \mathrm{d}z 
    = \int_\Omega d^i_l v_i  \, \mathrm{d}s \mathrm{d}z \,.
\end{equation}

In the equation above,
\begin{equation}
    \gamma_{\alpha \beta}(\mathbf{v}) = 
    \begin{bmatrix}
        v_3 / \lambda + \partial_s v_s & \left( \partial_s v_z + \partial_z v_s \right) / 2 \\[1mm]
        \left( \partial_s v_z + \partial_z v_s \right) / 2 & \partial_z v_z
    \end{bmatrix} ,
    \quad\,\, \rho_{\alpha \beta}(\mathbf{v}) =
    \begin{bmatrix}
        \partial_s^2 v_3 - \left( 2 v_3 + \lambda \partial_z v_z + 3 \lambda \partial_s v_s \right) / \lambda^2  & \partial_{s z} v_3- \partial_z v_s / \lambda \\[1mm]
        \partial_{s z} v_3 - \partial_z v_s / \lambda  & \partial_z^2 v_3
    \end{bmatrix}
    \label{eq:cylLinStain}
\end{equation}
are the linearized change of metric and change of curvature tensors associated with a virtual configuration described by the virtual displacement $\mathbf{v} = v_i \mathring{\mathbf{a}}^i$, and 
\begin{alignat}{2}
    n^{ss} (\mathbf{u}^l) & =  \frac{1}{1-\nu^2} \left( \gamma_{ss} (\mathbf{u}^l) + \nu \gamma_{zz} (\mathbf{u}^l)\right) , 
    \qquad & m^{ss} (\mathbf{u}^l)  &=  \frac{\eta^2}{12(1-\nu^2)} \left[ \left(\rho_{ss} (\mathbf{u}^l) + \kappa^l \right) + \nu \left( \rho_{zz} (\mathbf{u}^l) + \kappa^l \right) \right] , 
    \label{eq:cylLinStress1} \\
    n^{zz} (\mathbf{u}^l) & =  \frac{1}{1-\nu^2} \left( \gamma_{zz} (\mathbf{u}^l) + \nu \gamma_{ss} (\mathbf{u}^l)\right) ,
    \qquad & m^{zz} (\mathbf{u}^l) & =  \frac{\eta^2}{12(1-\nu^2)} \left[ \left(\rho_{zz} (\mathbf{u}^l) + \kappa^l \right) + \nu \left( \rho_{ss} (\mathbf{u}^l) + \kappa^l \right) \right] , 
    \label{eq:cylLinStress2} \\
    n^{sz} (\mathbf{u}^l )& = \frac{1}{1+\nu} \gamma_{sz} (\mathbf{u}^l),  
    \qquad & m^{sz} (\mathbf{u}^l) & = \frac{\eta^2}{12(1+\nu)} \rho_{sz} (\mathbf{u}^l) ,
    \label{eq:cylLinStress3}
\end{alignat}
are the linearized internal forces and moments associated with the displacement $\mathbf{u}^l$.
The linearized viscous force is
\begin{equation}
   \mathbf{d}_l = 
   d^i_l \, \mathring{\mathbf{a}}_i = 
   - \zeta_s \, \partial_t u_s^l \, \mathring{\mathbf{a}}_s - \zeta_z \,\partial_t u_z^l \, \mathring{\mathbf{a}}_z - \zeta_3 \, \partial_t u_3^l \, \mathring{\mathbf{a}}_3 ,
\end{equation}
whereas the linearized counterpart of the evolution equation~\eqref{eq:evLawNonDim} reads
\begin{equation}
\label{eq:evLawNonDimLinCyl}
   \partial_t \kappa^l + \kappa^l = \chi \,\partial_z u_3^l .
\end{equation}

The weak form~\eqref{eq:PVWExpNonDimLinCyl} is tested with $\mathbf{v} = v_s \, \mathring{\mathbf{a}}_s$, $\mathbf{v} = v_z \, \mathring{\mathbf{a}}_z$, and $\mathbf{v} = v_3 \, \mathring{\mathbf{a}}_3$, thereby reducing it to the three equations
\begin{align}
   \int_\Omega \bigg\{ n^{ss} (\mathbf{u}^l)\, \partial_s v_s + n^{s z} (\mathbf{u}^l)\, \partial_z v_s - m^{ss} (\mathbf{u}^l) \, \frac{3}{\lambda} \partial_s v_s  - 2 m^{s z} (\mathbf{u}^l) \, \frac{1}{\lambda} \partial_z v_s  + \frac{\zeta_s \xi}{\lambda \eta} \partial_t u^l_s \, v_s \bigg\}  \, \mathrm{d}s \mathrm{d}z &= 0 , 
   \label{eq:Balance1CylLin} \\[2mm]
   \int_\Omega \bigg\{ n^{zz} (\mathbf{u}^l) \, \partial_z v_z + n^{s z} (\mathbf{u}^l) \, \partial_s v_z - m^{ss} (\mathbf{u}^l) \,\frac{1}{\lambda} \partial_z v_z + \frac{\zeta_z \xi}{\lambda \eta} \partial_t u^l_z \, v_z \bigg\}  \, \mathrm{d}s \mathrm{d}z &= 0 , 
   \label{eq:Balance2CylLin} \\[2mm]
   \int_\Omega \bigg\{ n^{ss} (\mathbf{u}^l) \,\frac{1}{\lambda} v_3 + m^{ss} (\mathbf{u}^l) \,\bigg( \partial_s^2 v_3 - \frac{2}{\lambda^2} v_3 \bigg) + m^{zz} (\mathbf{u}^l) \,\partial_z^2 v_3 + 2 m^{s z} (\mathbf{u}^l) \, \partial_{s z} v_3 + \frac{\zeta_3 \xi}{\lambda \eta} \partial_t u^l_3 \, v_3 \bigg\}  \, \mathrm{d}s \mathrm{d}z &= 0 ,
   \label{eq:Balance3CylLin}
\end{align}
holding for any suitable test functions $v_s, \, v_z, \, v_3$.
In contrast to the linearized equations for plates~\eqref{eq:BalanceXPlateLin}--\eqref{eq:Balance3PlateLin}, for a cylindrical shell the bending and stretching problems are always coupled.
Equations~\eqref{eq:Balance1CylLin}--\eqref{eq:Balance3CylLin} are complemented by the weak form of the evolution law~\eqref{eq:evLawNonDimLinCyl}, namely
\begin{equation}
\label{eq:evLawNonDimLinCylWeak}
    \int_\Omega \bigg\{\left[ \partial_t \kappa^l + \kappa^l - \chi \, \partial_z u_3^l \right] \psi \bigg\}\,  \mathrm{d}s \mathrm{d}z = 0 ,
\end{equation}
where $\psi$ is the test function for the active curvature.

To proceed with the stability analysis, a dimensional reduction of Eqs.~\eqref{eq:Balance1CylLin}--\eqref{eq:evLawNonDimLinCylWeak} is carried out by exploiting the periodicity of all fields in the coordinate $s \in [0,2\pi\lambda]$. In particular, the components $u_i^l$ of the displacement field, with $i = \{s,z,3\}$, and the spontaneous curvature $\kappa^l$ are expanded in Fourier series as
\begin{align}
    u_i^l (s,z,t) &= a_i^0 (z,t) + \sum_{m=1}^{+\infty} \left\{ a_i^m (z,t) \cos{\left( m \frac{s}{\lambda}\right)} + b_i^m (z,t) \sin{\left( m \frac{s}{\lambda}\right)} \right\} ,
    \label{eq:LinCylFourierDispl} \\
    \kappa^l (s,z,t) &= a_\kappa^0 (z,t) + \sum_{m=1}^{+\infty} \left\{ a_\kappa^m (z,t) \cos{\left( m \frac{s}{\lambda}\right)} + b_\kappa^m (z,t) \sin{\left( m \frac{s}{\lambda}\right)} \right\} ,
    \label{eq:LinCylFourierK}
\end{align}
where $\{a^m_s,b^m_s,a^m_z,b^m_z,a^m_3,b^m_3,a^m_\kappa,b^m_\kappa\}$ are  unknown functions to be determined for every integer $m \geq 0$. The Fourier series expansion is adopted also for the relevant test functions $v_i$, with $i = \{s,z,3\}$, and $\psi$, leading to  
\begin{align}
    v_i (s,z) &= \alpha_i^0 (z) + \sum_{m=1}^{+\infty} \left\{ \alpha_i^m (z) \cos{\left( m \frac{s}{\lambda}\right)} + \beta_i^m (z) \sin{\left( m \frac{s}{\lambda}\right)} \right\} ,
    \label{eq:LinCylFourierDisplTest} \\
    \psi (s,z) &= \alpha_\kappa^0 (z) + \sum_{m=1}^{+\infty} \left\{ \alpha_\kappa^m (z) \cos{\left( m \frac{s}{\lambda}\right)} + \beta_\kappa^m (z) \sin{\left( m \frac{s}{\lambda}\right)} \right\} .
    \label{eq:LinCylFourierKTest}
\end{align}
Note that, in the expansions above, the superscript $m$ is a mode index and not an exponent. By virtue of the orthogonality of the sine and cosine functions, substitution of the Fourier series~\eqref{eq:LinCylFourierDispl}--\eqref{eq:LinCylFourierKTest} into~\eqref{eq:Balance1CylLin}--\eqref{eq:evLawNonDimLinCylWeak} allows for the explicit integration in the coordinate $s$. 
Remarkably, upon fixing $n \in \mathbb{N}$ and testing the resulting weak equations with
\begin{equation}
    \alpha_i^m =
    \begin{cases}
       \alpha_i & \text{for }\,m = n \\
       0 & \text{for }\,m \neq n
    \end{cases}\quad \text{and}
    \quad
    \beta_i^m = 0,
\end{equation}
or with
\begin{equation}
    \alpha_i^m = 0 \quad \text{and}
    \quad
    \beta_i^m = 
        \begin{cases}
       \beta_i & \text{for }\,m = n \\
       0 & \text{for }\,m \neq n
    \end{cases} ,
\end{equation}
for every integer $m \geq 0$ and $i = \{s,z,3,\kappa\}$, we obtain an infinite set of decoupled, one-dimensional systems (indexed by $n$) of eight equations, in the unknowns $\{a^n_s,b^n_s,a^n_z,b^n_z,a^n_3,b^n_3,a^n_\kappa,b^n_\kappa\}$.
A closer inspection reveals that each such system splits into two identical, decoupled systems of four equations, in the unknowns $\{b^n_s,a^n_z,a^n_3,a^n_\kappa\}$ or, equivalently, $\{-a^n_s,b^n_z,b^n_3,b^n_\kappa\}$.
In general, this would not imply equality of the corresponding unknowns, since different initial conditions may yield different solutions of the same equations.
However, since we are concerned with the stability of the trivial equilibrium, we solve an eigenvalue problem rather than an initial-value problem for the linearized equations. Therefore, we assume
\begin{equation} \label{eq:LinCyl1Dsimpl}
    a^n_s = - b^n_s, \qquad 
    b^n_z = a^n_z, \qquad
    b^n_3 = a^n_3, \qquad
    b^n_\kappa = a^n_\kappa,
\end{equation}
without loss of generality.

In summary, we have reduced the dimensionality of the problem, passing from a system of four equations in two spatial dimensions to an infinite set of decoupled systems of four equations in the single spatial variable $z$, which we report here for completeness:
\begin{gather}
    \begin{split}
        \int_0^1 \bigg\{ \bigg[ \nu \, \partial_z^2 a_3^n - \bigg( \frac{12 \nu}{\eta^2} + \frac{2 + n^2}{\lambda^2} \bigg) \, a_3^n + ( 1+\nu) \, a_\kappa^n - \bigg(\frac{12 n \nu}{\eta^2} + \frac{3 n}{\lambda^2} \bigg) \, b_s^n - \bigg(\frac{12 \lambda}{\eta^2} + \frac{1}{\lambda} \bigg) \, \partial_z a_z^n  \bigg]  \partial_z \alpha_z \,&- \\[1.5mm]
         6 \,(1-\nu)\, \bigg[ \frac{2 \xi \zeta_z (1+\nu)}{\eta^3} \, \partial_t a_z^n + \frac{n}{\lambda \eta^2} \Big(n a_z^n - \lambda \partial_z b_s^n \Big) \bigg] \, \alpha_z \bigg\}  \, \mathrm{d}z &= 0\,,  \quad \forall \, \alpha_z\,,
    \end{split} \label{eq:Balance1CylLin1D} \\[4.5mm]
    \begin{split}
        \int_0^1 \bigg\{ \bigg[ - 2 (1-\nu) \Big(n \partial_z a_3^n + \partial_z b_s^n \Big) - 6 (1-\nu) \frac{\lambda}{\eta^2} \Big(\lambda \partial_z b_s^n - n a_z^n \Big) \bigg] \, \partial_z \beta_s \,&+ \\[1.5mm]
        \bigg[ - \frac{12 \xi \zeta_s \lambda (1-\nu^2)}{\eta^3} \, \partial_t b_s^n - \frac{12 n}{\eta^2} \Big(a_3^n + n b_s^n + \lambda \nu \partial_s a_z^n\Big) \,&+ \\[1.5mm]
        3 n \bigg[\nu \,\partial_z^2 a_3^n - \frac{2+n^2}{\lambda^2} a_3^n - \frac{3 n}{\lambda^2} b_s^n + (1+\nu) a_\kappa^n - \frac{\partial_z a_z^n}{\lambda} \bigg] \bigg] \, \beta_s \bigg\}  \, \mathrm{d}z &= 0\,, \quad \forall \, \beta_s,
    \end{split} \label{eq:Balance2CylLin1D} \\[4.5mm]
    \begin{split}
        \int_0^1 \bigg\{ \bigg[ \frac{(2+n^2) \nu }{\lambda^2} a_3^n - (1+\nu) a_\kappa^n + \frac{3 n \nu}{\lambda^2} b_s^n + \frac{\nu}{\lambda} \partial_z a_z^n - \partial_z^2 a_3^n \bigg] \, \partial_z^2 \alpha_3 - \bigg[\frac{2 n}{\lambda^2} (1-\nu) \Big(n \partial_z a_3^n + \partial_z b_s^n\Big) \bigg]  \, \partial_z \alpha_3 &\,+\\[1.5mm]
        \bigg[\bigg[ \frac{2+n^2}{\lambda^2} \nu \partial_z^2 a_3^n - \frac{12 \xi \zeta_3 (1-\nu^2)}{\lambda \eta^3} \partial_t a_3^n  - \bigg( \frac{12 \nu}{\lambda \eta^2} + \frac{2+n^2}{\lambda^3} \bigg)\, \partial_z a_z^n \bigg] &\,+ \\[1.5mm]
        \frac{(2+n^2)(1+\nu)}{\lambda^2} a_\kappa^n - 3 n \bigg(\frac{4}{\lambda^2 \eta^2} + \frac{2+n^2}{\lambda^4} \bigg) b_s^n - a_3^n \bigg( \frac{12}{\lambda^2 \eta^2} + \frac{(2+n^2)^2}{\lambda^4} \bigg) \bigg]  \, \alpha_3 \bigg\}  \, \mathrm{d}z &= 0\,, \quad \forall \,\alpha_3,
    \end{split}  \label{eq:Balance3CylLin1D} \\[4.5mm]
    \int_0^1 \bigg\{ \left[ \partial_t a_\kappa^n + a_\kappa^n - \chi \, \partial_z a_3^n \right] \alpha_\kappa \bigg\}  \, \mathrm{d}z = 0, \quad \forall \,\alpha_\kappa.   \label{eq:EvLawCylLin1D}
\end{gather}

We remark that the dimensional reduction has been carried out on the weak form of the linearized governing equations, since their numerical solution relies on the finite element method.
Analogous manipulations can be used to reduce the dimensionality of the problem when the governing equations are first recast in strong form.
Seeking time-harmonic solutions of the form
\begin{equation} \label{eq:LinCyl1DtimeHarmonic}
    b_s^n(z,t) = \mathrm{exp} (\omega_n \, t) \, \hat{b}_s^n (z) , \quad
    a_z^n(z,t) = \mathrm{exp} (\omega_n \, t) \, \hat{a}_z^n (z) , \quad
    a_3^n(z,t) = \mathrm{exp} (\omega_n \, t) \, \hat{a}_3^n (z) , \quad
    a_\kappa^n(z,t) = \mathrm{exp} (\omega_n \, t) \, \hat{a}_\kappa^n (z) ,
\end{equation}
we obtain the corresponding eigenvalue problem, which we solve by the finite element method, as detailed in~\ref{sec:appDetailsCylindersStab}, to characterize the complex frequency $\omega_n$ and the associated eigenfunctions for each value of the circumferential mode $n$.
In practice, we fix an integer $M$ and solve the $M+1$ eigenvalue problems corresponding to $n=0,\dots,M$.
Here, we choose $M=20$.
We denote by $\omega_{\mathrm{max}} = \alpha \pm \mathrm{i} \, \beta$ the eigenvalue with the largest real part among all the $M+1$ eigenvalue problems, and by $n_\mathrm{max}$ the corresponding circumferential mode.

We report in Fig.~\ref{fig:CylGeometry}b the stability maps for $\nu = 0.3$, $\eta = 1.0 \times 10^{-3}$ and $\xi = \{ \xi_\mathrm{ref}, 5 \xi_\mathrm{ref}, 20 \xi_\mathrm{ref} \}$, with $\xi_\mathrm{ref} = 2.967 \times 10^{-8}$ the value used in Section~\ref{sec:plates} (see~\ref{sec:appElasticPEH}), as functions of the dimensionless target curvature $\chi \in [0,1400]$ and the cylinder aspect ratio $\lambda \in [0.05, 0.5]$.
We identify the parameter regions in which the system is stable (blue), and color-code the regions in which it exhibits flutter instability (from yellow to green) according to the circumferential mode $n_\mathrm{max}$ that maximizes the real part $\alpha$ of the eigenvalue, which can be predicted as a function of the model parameters.
Within the parameter range investigated, no flutter occurs for $n=0$ and $n=1$. We observe that $n_\mathrm{max}$ increases with the cylinder radius $\lambda$.
Furthermore, as $\xi$ increases, flutter regions associated with higher values of $n_\mathrm{max}$ emerge, up to the value of $n_\mathrm{max} = 14$. 
In addition to the stability regions, we show in Fig.~\ref{fig:CylGeometry}c representative shapes of the cylindrical shell corresponding to the most unstable flutter eigenmodes. These are computed from
\begin{gather}
    \hat{u}_s^n (s,z) =  b_s^n (z) \, \Big[ \sin{\left( n \frac{s}{\lambda}\right)} - \cos{\left( n \frac{s}{\lambda}\right)} \Big], \\[1mm]
    \hat{u}_z^n (s,z) =  a_z^n (z) \, \Big[ \cos{\left( n \frac{s}{\lambda}\right)} + \sin{\left( n \frac{s}{\lambda}\right)} \Big], \\[1mm]
    \hat{u}_3^n (s,z) =  a_3^n (z) \, \Big[ \cos{\left( n \frac{s}{\lambda}\right)} + \sin{\left( n \frac{s}{\lambda}\right)} \Big], 
\end{gather}
which give the displacement components of the eigenmode with $n = n_\mathrm{max}$, obtained by combining relations~\eqref{eq:LinCylFourierDispl}, \eqref{eq:LinCyl1Dsimpl}, and~\eqref{eq:LinCyl1DtimeHarmonic}.
All modes are evaluated at the representative value of $\chi = 1200$. The figure reports both the real and the imaginary part of the eigenmode, together with a top view of a cross-section of the deformed cylinder at half-height ($z = 1/2$) and the computed flutter eigenvalue $\omega_\mathrm{max}$. In particular,
the first row shows the modes for $\xi = \xi_\mathrm{ref}$ and $\lambda = \{ 0.2, 0.35, 0.5 \}$, corresponding to $n_\mathrm{max} = \{ 2,3,4 \}$, respectively. These illustrate how the real and imaginary parts of $\omega_\mathrm{max}$ increase with the radius.
Similarly, the second row shows the modes for $\xi = 20\, \xi_\mathrm{ref}$ and $\lambda = \{ 0.15, 0.2, 0.23 \}$, corresponding to $n_\mathrm{max} = \{ 5,6,7 \}$, respectively.
In these conditions, a smaller change in $\lambda$ suffices to increase $n_\mathrm{max}$, consistently with the right panel of Fig.~\ref{fig:CylGeometry}b. Clearly, as $n_\mathrm{max}$ increases, the eigenmodes exhibit an increasing number of circumferential lobes and a progressively shorter axial wavelength. As for the case of elliptic plates, the real and imaginary parts of the eigenmodes are not proportional, such that their modulation in time provides nonreciprocal shape changes.

%%%%%%
\subsubsection{Unstretchable cylinders cannot flutter}
\label{sec:cylindricalShells:stability:unstretc}

In Section~\ref{sec:plates:stability}, it was shown that the linearized problem for active plates is insensitive to stretching. By contrast, in Section~\ref{sec:cylindricalShells:stability}, stretching and bending were shown to remain coupled for a cylindrical shell in the linear setting. It is therefore of interest to explore whether enforcing shell inextensibility merely restricts the class of unstable solutions or, more fundamentally, whether it completely suppresses the instability.
To this end, we enforce such kinematic constraint on the trial and test functions, adapting the computations of~\citet{timosenkoTheoryPlatesShells1996}. In particular,
we impose vanishing in-plane strains, recall equation~\eqref{eq:cylLinStain}$_1$, by requiring
\begin{equation}
    \dfrac{u_3^l}{\lambda} + \partial_s u_s^l =  0 \,, \qquad
    \dfrac{1}{2} \left( \partial_s u_z^l + \partial_z u_s^l \right) = 0 \,, \qquad 
    \partial_z u_z^l = 0 . 
\end{equation}

By exploiting the periodicity condition in $s$, the displacement components take the form
\begin{align}
    u_z^l (s,t) &= a_z^0(t) +  \sum_{n=1}^{+\infty} \left\{ a_z^n (t) \cos{\left( n \frac{s}{\lambda}\right)} + b_z^n (t) \sin{\left( n \frac{s}{\lambda}\right)} \right\} ,
    \label{eq:troppe1} \\
    u_s^l (s,z,t) &= a_s^0(t) +  \sum_{n=1}^{+\infty} \left\{ a_s^n(t) \cos{\left( n \frac{s}{\lambda}\right)} + b_s^n(t) \sin{\left( n \frac{s}{\lambda}\right)} - \frac{z}{\lambda}  n \left[ b_z^n (t) \cos{\left( n \frac{s}{\lambda}\right)} - a_z^n (t) \sin{\left( n \frac{s}{\lambda}\right)} \right] \right\} ,
    \label{eq:troppe2}\\
    u_3^l (s,z,t) &= - \sum_{n=1}^{+\infty} \left\{ n \left[ b_s^n (t) \cos{\left( n \frac{s}{\lambda}\right)} - a_s^n (t) \sin{\left( n \frac{s}{\lambda}\right)} \right] + \frac{z}{\lambda} n^2 \left[ a_z^n (t) \cos{\left( n \frac{s}{\lambda}\right)} + b_z^n (t) \sin{\left( n \frac{s}{\lambda}\right)} \right] \right\} .
    \label{eq:troppe3}
\end{align}
In particular, we notice that $u_z^l$ is independent of $z$, whereas $u_s^l$ and $u_3^l$ are linear in $z$. Hence, $\partial_z u_3^l$ is independent of $z$ and, by equation~\eqref{eq:evLawNonDimLinCyl}, so is $\kappa^l$, namely
\begin{equation}
    \kappa^l (s,t) = a_\kappa^0(t) +  \sum_{n=1}^{+\infty} \left\{ a_\kappa^n(t) \cos{\left( n \frac{s}{\lambda}\right)} + b_\kappa^n(t) \sin{\left( n \frac{s}{\lambda}\right)} \right\} .
\end{equation}

Inserting these series expansions into~\eqref{eq:PVWExpNonDimLinCyl} and~\eqref{eq:evLawNonDimLinCylWeak}, with similar expressions for the admissible test functions $\{ v_s,v_z,v_3,\psi \}$, allows the integrations in both $s$ and $z$ to be carried out explicitly so as to obtain the strong form of the governing equations:
\begin{gather}
\begin{split}
    n^2 \Big(n^2-1\Big) \eta^3 \left\{\Big(n^2-1\Big) \left[2 \Big(n^2+6\lambda^2(1-\nu)\Big)\, a_z^n + 3 n \lambda b_s^n \right] + 3 \lambda^3 (1+\nu) \,a_\kappa^n \right\} \,&+  \\[1.5mm]
    12 \xi \lambda^3 \Big(1 - \nu^2\Big) \left[ 2 \Big(n^4 \zeta_3 + 3 \lambda^2 \zeta_z + n^2 \zeta_s\Big) \,\partial_t a_z^n + 3 n \lambda \,\Big(n^2 \zeta_3 + \zeta_s\Big)\, \partial_t b_s^n \right] &= 0 \,,
\end{split} 
\label{eq:Balance1CylLin1DUnstr} \\[3mm]
n\Big(n^2-1\Big) \eta^3 \left[n\Big(n^2-1\Big)\Big(n a_z^n + 2 \lambda b_s^n\Big) + 2 \lambda^3 (1+\nu) a_\kappa^n \right] + 12 \xi \lambda^3 \Big(1 - \nu^2\Big) \Big(n^2 \zeta_3 + \zeta_s\Big) \Big(n \,\partial_t a_z^n + 2 \lambda \,\partial_t b_s^n\Big) = 0 \,,
\label{eq:Balance2CylLin1DUnstr} \\[3mm]
\partial_t a_\kappa^n + a_\kappa^n + \chi \,\frac{n^2}{\lambda} a_z^n = 0 \,.
\label{eq:Balance3CylLin1DUnstr}
\end{gather}
These constitute a system of three decoupled ordinary differential equations in time, in the unknowns $\{ a_z^n, b_s^n, a_\kappa^n \}$ for every integer $n \geq 0$. We remark that, as in Section~\ref{sec:cylindricalShells:stability}, the computations again yield $b_z^n = a_z^n$, $a_s^n = - b_s^n$, and $b_\kappa^n = a_\kappa^n$. Under the time-harmonic assumption of
\begin{equation}
    a_z^n(t) = \mathrm{exp} (\omega_n \, t) \, \hat{a}_z^n , \quad 
    b_s^n(t) = \mathrm{exp} (\omega_n \, t) \, \hat{b}_s^n , \quad 
    a_\kappa^n(t) = \mathrm{exp} (\omega_n \, t) \, \hat{a}_\kappa^n , 
\end{equation}
we obtain the algebraic eigenvalue problem associated with~\eqref{eq:Balance1CylLin1DUnstr}--\eqref{eq:Balance3CylLin1DUnstr}, whose characteristic equation admits the explicit solutions
\begin{gather}
    \omega_{n,1} = -1 ,\qquad 
    \omega_{n,2} = - \frac{\eta^3 n^2 \big(n^2-1\big)^2 }{12 \xi \lambda^3 \big( n^2 \zeta_3 + \zeta_s \big)\big(1-\nu^2\big)} ,\qquad 
    \omega_{n,3} = - \frac{\eta^3 n^2 \big(n^2-1\big)^2 \big[n^2 + 24 \lambda^2 (1-\nu) \big]}{12 \xi \lambda^3 \big[ 12 \lambda^2 \zeta_z + n^2 \big( n^2 \zeta_3 + \zeta_s \big) \big(1-\nu^2\big) \big]} ,
\end{gather}
for every integer $n \geq 0$.
We note that the eigenvalues are always real and non-positive, vanishing only for $n=0$ and $n=1$, and independent of the loading parameter $\chi$. Hence, no propagating eigenmode can arise and unstretchable cylindrical shells can only relax monotonically toward the trivial equilibrium once perturbed, undergoing viscous relaxation in the surrounding fluid.
This is a consequence of the limited set of shapes that an unstretchable shell can assume: at the linear level, they comprise only inextensional cross-sectional deformations (ovalization and higher-order lobed modes), either constant or varying linearly along $z$, according to~\eqref{eq:troppe1}--\eqref{eq:troppe3}.
Such deformations cannot sustain the nonreciprocal traveling waves that are characteristic of flutter instability, as shown in Section~\ref{sec:plates:simulations} for plates.
We therefore expect the nonlinear simulations to reveal that, unlike for plates, a certain amount of stretching is necessary for the system to sustain flutter, even though it is energetically costly compared to bending.

\subsection{Nonlinear numerical results}
\label{sec:cylindricalShells:simulations}

\begin{figure}[p]
	\centering
	\includegraphics[width=\linewidth]{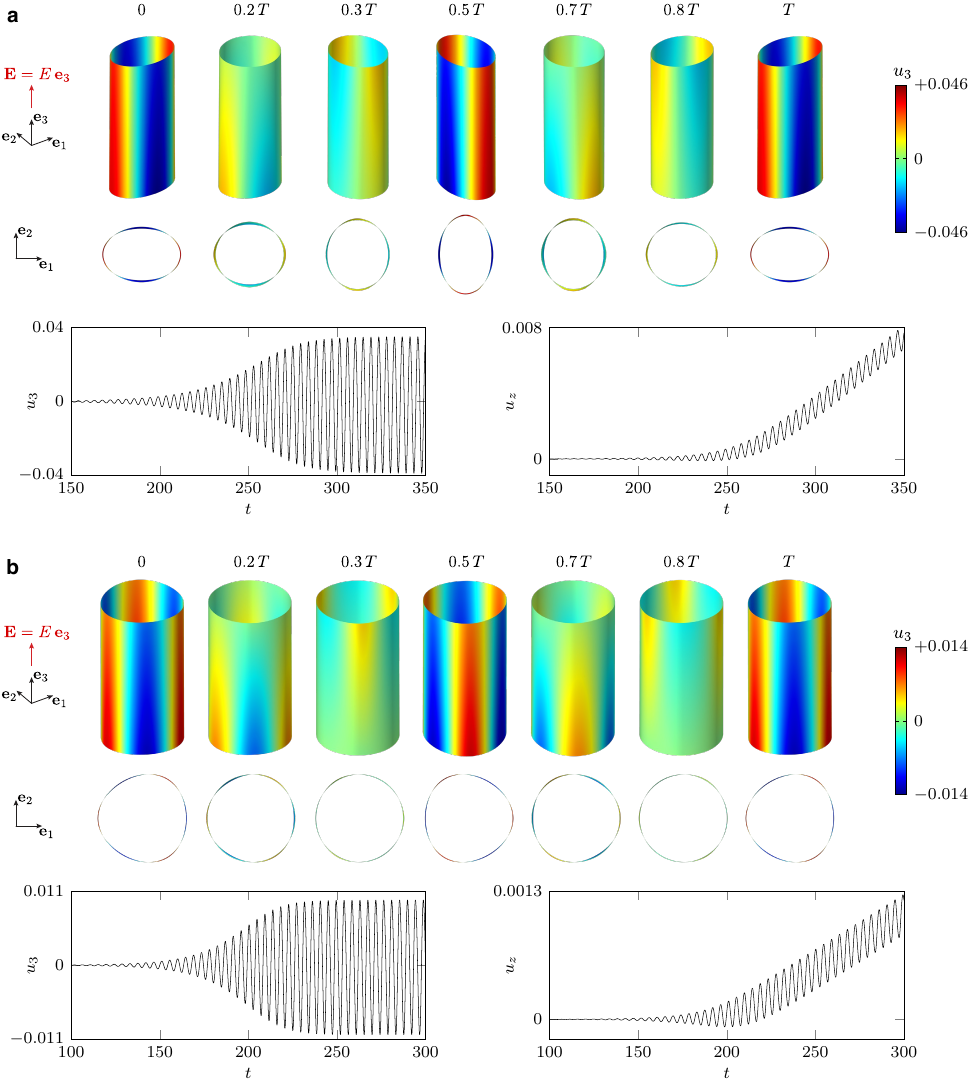}
	\caption{
		Numerical simulations of the nonlinear model for  cylindrical shells of circular cross-section with $\nu = 0.3$ and $\eta = 1.0 \times 10^{-3}$.
		Panel (a) corresponds to $\chi = 700$, $\lambda = 0.2$, and $\xi = \xi_\mathrm{ref}$, while panel (b) corresponds to $\chi = 800$, $\lambda = 0.27$, and $\xi = 5\, \xi_\mathrm{ref}$.
		In both panels, the top row displays seven snapshots of the deformed configuration covering one period $T$, once the limit cycle is reached, from a three-dimensional view and from a top view.
		The bottom row shows the time evolution of the transverse (left) and axial (right) displacements of the material point corresponding to $(\pi \lambda, 0.5) \in \Omega$.
        See also Supplementary Video~4.
	}
	\label{fig:CylNumericalResults}
\end{figure}

We conclude our analysis of active cylindrical shells by performing numerical simulations of the nonlinear model; details of the numerical implementation are provided in~\ref{sec:appDetailsCylindersNonlin}.
We report in Fig.~\ref{fig:CylNumericalResults} the results of simulations performed with $\nu = 0.3$ and $\eta = 1.0 \times 10^{-3}$, see also Supplementary Video~4.
In particular, Fig.~\ref{fig:CylNumericalResults}a shows the results for $\chi = 700$, $\lambda = 0.2$, and $\xi = \xi_\mathrm{ref} = 2.967 \times 10^{-8}$, a choice of parameters for which the linear stability analysis predicts flutter with $n_\mathrm{max} = 2$, as reported in the left panel of Fig.~\ref{fig:CylGeometry}b, and $\omega_\mathrm{max} = 0.04 \pm \mathrm{i} \, 1.38$.
In parallel, Fig.~\ref{fig:CylNumericalResults}b shows results for $\chi = 800$, $\lambda = 0.27$, and $\xi = 5\, \xi_\mathrm{ref}$, for which the linear stability analysis predicts flutter with $n_\mathrm{max} = 3$, as reported in the central panel of Fig.~\ref{fig:CylGeometry}b, and $\omega_\mathrm{max} = 0.06 \pm \mathrm{i} \, 1.39$.
The two simulations yield qualitatively analogous behavior. 
Following a perturbation of the trivial equilibrium, the deformation of the cylindrical shell approaches the leading eigenmode predicted by the stability analysis after a short transient. The displacement amplitude grows while the shell oscillates periodically, exhibiting a traveling wave propagating in the direction opposite to that of the electric field. The system eventually settles onto a limit cycle, consistently with a Hopf bifurcation; the prolonged transient preceding the limit cycle is consistent with the real part of $\omega_\mathrm{max}$, which sets the growth rate in the small-amplitude regime.
The period of the resulting oscillation, $T \approx 4.41$ for Fig.~\ref{fig:CylNumericalResults}a and $T \approx 4.53$ for Fig.~\ref{fig:CylNumericalResults}b, is in remarkable agreement with the value $2\pi/\beta$ predicted by the imaginary part $\beta$ of the leading eigenvalue, which is 4.55 and 4.52, respectively, confirming that the nonlinear dynamics remains organized around the leading eigenmode.
The top row of each panel displays seven shapes sampled within one period of the established limit cycle, illustrating the distinctive nature of the self-sustained oscillatory motion. At the beginning and midpoint of the deformation cycle, the configurations are almost pure cross-sectional deformations of the reference shape, which require no stretching. Between these instants, a rapid wave travels along the cylinder, so that stretching, which is indispensable for the onset of flutter as established in Section~\ref{sec:cylindricalShells:stability:unstretc}, is concentrated within this short time interval.
Although if less pronounced than for plates, cylinders also swim in the direction of the electric field, as shown in the bottom-right plots of each panel. The magnitude of the net displacement is consistent with the proximity of the loading parameter to the flutter threshold: close to the instability onset, the stretching that underlies the nonreciprocity of the motion remains limited, and with it the net motion it determined. A quantitative refinement of this estimate would also require a more accurate characterization of the resistive coefficients employed in the numerical simulations, whose present approximation may underestimate the viscous interactions.

%%%%%%%%%%%%%%%%%%%%%%%%%%%%%%%%%%%%%%%%%%%%%%%%%%%%%%%%%%%%%%%%

\section{Experiments on PEH samples}
\label{sec:experiments}

In this section, we confront the theoretical predictions of Section~\ref{sec:plates} with experiments carried out on PEH elliptic plates, fabricated with the aspect ratio of $\lambda = 2/3$ considered in the simulations reported in Fig.~\ref{fig:PlatesNonlinearResults}c.

\subsection{Fabrication of hydrogel samples}
\label{sec:exp_fab}

The samples were made from \emph{poly-acrylamide-co-sodium acrylate} (PAAm-co-SA) starting from a pre-gel solution with reagent concentrations as in Table~\ref{tab:stk_cnc}. 
\begin{table}[t]
	\centering
	\footnotesize
	\caption{Pre-gel solution composition from stock solutions. The sodium acrylate and the photoinitiator VA-086 were prepared from powder and diluted with deionized water, while acrylamide and bis-acrylamide were supplied as ready-to-use solutions.}
		\begin{tabular}{l c}
			\toprule
			Stock solution & Quantity [\si{\micro\liter}] \\
			\hline
			Acrylamide (40\% w/v), Fisher Bioreagents n. BP1402-1 & 287 \\
			Bis-acrylamide (2\% w/v), Fisher Bioreagents n. BP1404-250 & 263 \\
			Sodium acrylate (10\% w/v), Sigma-Aldrich n. 408220-25G & 239\\
			VA-086 (3\% w/v), Wako Chemicals n. 013-19342 & 168 \\
			Fluorescent microspheres (2\% w/v) \SI{1.0}{\micro\meter}, Invitrogen n. F13081 &
			43 \\
			\bottomrule
	\end{tabular}
	\label{tab:stk_cnc}
\end{table}
The diameter and concentration of the fluorescent beads were carefully chosen to obtain a speckle pattern suitable for the application of digital image correlation (DIC) techniques to reconstruct the sample motion, as later  detailed. The optimal concentration was identified after several trials to ensure both adequate resolution of the speckle pattern and light emission intensity.

\begin{figure}[t]
	\centering
	\includegraphics[width=\textwidth]{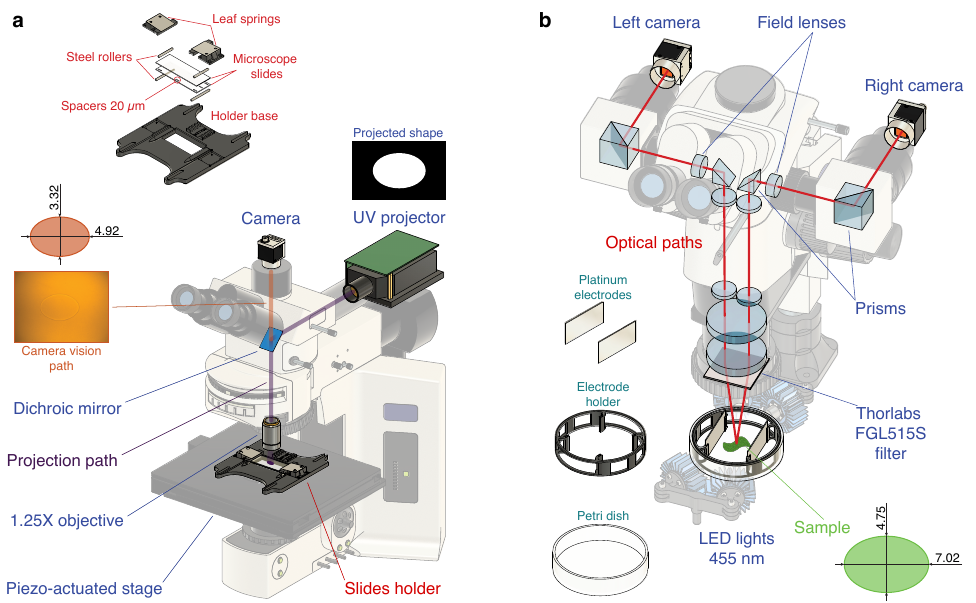}
	\caption{Fabrication and experimental setups. (a) Fabrication setup based on the upright Olympus BX61 microscope. The microscope is equipped with the piezo-actuated stage BX3-SSU and modified to use the UV DLP projector Pro 4500 from Wintech. A dichroic mirror is employed to project UV light through the $1.25\times$ objective to the stage. The Basler acA4024-29uc camera is employed to monitor the polymerization process. The slides containing the photoinitiated pre-gel solution are held in position by means of a custom-made holder, whose exploded representation is shown in the upper part of the panel. (b) Experimental setup for 3D-DIC measurement of the PEH plate. The stereo optical microscope Olympus SZX16 is equipped with two Basler acA1920-155um cameras to perform stereo imaging of the sample. The small-scale electrolytic cell is realized using a Petri dish equipped with platinum electrodes, which are held in position by a custom-made, 3D-printed holder that ensures a \SI{45}{\milli\meter} distance between the electrodes. The setup is designed to exploit fluorescence by combining the illumination, provided by four LEDs of wavelength \SI{455}{\nano\meter}, with an optical filter having a cut-off wavelength of \SI{515}{\nano\meter}. The measured dimensions of the sample after the fabrication process are reported in panel (a), while panel (b) shows its dimensions under the test condition, namely after the sample has swollen in a \SI{1}{\gram\per\liter} solution of NaCl in water.}
	\label{fig:general_exp}
\end{figure}

The fabrication protocol consisted of two steps: the preparation of the pre-gel solution and the subsequent polymerization process. In the first step, all the reagents, except the photoinitiator, were mixed and degassed for twenty minutes at a pressure of \SI{0.2}{\bar}. The photoinitiator was then added to the solution immediately before polymerization.
As for the second step, polymerization was performed with a custom-made photolithography setup, based on an upright microscope (Olympus BX61) as shown in Fig.~\ref{fig:general_exp}a.
The elliptical pattern was projected from a UV DLP projector (Wintech DLP PRO 4500), emitting at a wavelength of \SI{385}{\nano\meter} with a power density of \SI{244.6}{\micro\watt\per\square\milli\meter}. The optical arrangement ensured that the ellipse projected onto the focal plane had major and minor axes of \SI{4.92}{\milli\meter} and \SI{3.32}{\milli\meter}, respectively, resembling the aspect ratio $\lambda = 2/3$ investigated in Section~\ref{sec:plates}. In particular, the UV beam was reflected downward to the microscope objective (Olympus PlanApo N $1.25\times$) by a dichroic mirror (Thorlabs Cold Mirror FM04R). The sample was imaged during fabrication using a digital camera (Basler acA4024-29uc) mounted on a $0.5\times$ adapter, which provided a field of view adequate to monitor the polymerization process.
A \SI{50}{\micro\liter} volume of photoinitiated pre-gel solution was injected between two soda-lime microscope slides, separated by \SI{20}{\micro\meter} via eight symmetrically arranged polymeric stripes. These polymeric spacers were printed on the bottom slide using a two-photon-polymerization system (3D~MF solution set of Photonic Professional GT2 from Nanoscribe GmbH) to ensure precise control of the ellipse thickness during fabrication, see Fig.~\ref{fig:general_exp}a. The slides were held in position by a custom-made holder equipped with two pairs of aligned steel rollers, preloaded by two leaf springs to ensure contact between the rollers and the slides. The slide holder was fixed to a piezo-actuated microscope stage (Olympus BX3-SSU) by four nylon fasteners.
Typically, a slide held 6--8 ellipses, each subjected to a two-minute UV exposure.

Following polymerization, the slides were immersed in deionized water and stored at \SI{5}{\celsius} until the ellipses detached naturally. Once detached, they were transferred to a Petri dish containing a \SI{1}{\gram\per\liter} solution of NaCl in water and stored at \SI{5}{\celsius}. After immersion in the saline bath, the samples reached a stable configuration with  major and minor axes of \SI{7.02}{\milli\meter} and \SI{4.75}{\milli\meter}, respectively. 
Assuming isotropic swelling, we estimated a swelling factor of \num{1.43}.
Before each experimental session, the hydrogel samples were thermally equilibrated at the room temperature of \SI{24}{\celsius} for at least twenty minutes.

\subsection{Experimental methods and setup}

The experiments consisted in stimulating the PEH  samples with a constant and uniform electric field within a small-scale electrolytic cell.
The main goal of the campaign was to induce flutter instability in the elliptic plates and to reconstruct the resulting swimming dynamics associated with thrust generation, as discussed in Section~\ref{sec:plates:simulations}.
This was achieved through 3D-DIC, an established experimental technique that exploits the synergy between 3D computer vision~\citep{Hartley2004-MVGCV}, used to reconstruct a surface starting from two or more planar views, and 2D-DIC, used to identify matching points across these views~\citep{Sutton2009-DIC}.
The successful application of this technique has been previously reported in several technical studies~\citep{Schreier2004,Zhang2006,Sutton2008-JBMRPA}, including the use of fluorescence to achieve speckle patterns suitable for DIC measurements.
The small size of the specimens required high-resolution image acquisition during the experiments, while the application of 3D-DIC necessitated a stereo imaging system. Both requirements were met using an optical stereo microscope equipped with a dual-camera imaging system.

The experimental apparatus is shown in Fig.~\ref{fig:general_exp}b. The core of the imaging system is an Olympus SZX16 stereo microscope, equipped with a SDFPLAPO1XPF objective and integrated with a SZX2-LBS beam splitter. This was used to intercept the optical paths of the microscope and to direct them towards two SZX-PHA optical ports, each incorporating a field lens and a reflecting prism to focus and deviate light onto a digital camera (Basler acA1920-155um) mounted on a $1\times$ U-TV1X-2 C-mount adapter. By this configuration, the SZX16 ensured high-resolution, digital stereo imaging, thereby enabling 3D-DIC measurements.
High-contrast imaging of the PEH samples was achieved through fluorescence microscopy. Specifically, the samples were illuminated by two pairs of LEDs with a nominal wavelength of \SI{455}{\nano\meter}, while the microscope objective was equipped with a long-pass filter (model FGL515S from Thorlabs) with cut-off wavelength of \SI{515}{\nano\meter}, see Fig.~\ref{fig:general_exp}b.
The electrolyte solution consisted in \SI{1}{\gram\per\liter} of NaCl in water, while the electrodes were two $50$\,\si{\milli\meter} $\times$ $25$\,\si{\milli\meter} platinum plates of thickness \SI{0.1}{\milli\meter}, connected to a stabilized power supply (Rohde \& Schwarz HMP4040), which provided a constant voltage difference across them. The electrodes were held parallel at a relative distance of \SI{45}{\milli\meter} by a custom-made holder within a Petri dish of diameter \SI{90}{\milli\meter}.

The calibration of the stereo imaging system
(SZX16 microscope and pair of acA1920-155um cameras)
was performed using the Stereo Camera Calibrator app of MATLAB (version R2024b) and a target with \num{189} circles of \SI{250}{\micro\meter} in diameter arranged on a $9\times21$ staggered grid with horizontal pitch of \SI{500}{\micro\meter} and half-pitch vertical spacing and row offset. The target was fabricated with the system Photonic Professional GT2 (Nanoscribe GmbH) using the 3D~MF solution set, and mounted on a gimbal device coupled with a two-axis micrometric stage, allowing for rigid rotations and translations of the target.
Specifically, \num{16} image pairs of the target were acquired at different orientations and positions, and then passed as input to the calibrator.
%which relies on the pinhole camera model along with a two-coefficient distortion model for radial aberrations.
The intrinsic parameters of each camera were preliminary estimated through the Single Camera Calibrator app of MATLAB. Before calibration, the microscope magnification was set to $0.8\,\times$, leading to a field of view of $14.32$\,\si{\milli\meter}\,$\times$\,$8.95$\,\si{\milli\meter} and a measured scale factor of \SI{7.457}{\micro\meter\per\pixel}.
\begin{table}[t]
	\centering
	\footnotesize
	\caption{Model parameters used to perform the 2D-DIC analysis on the elliptic PEH samples to identify matching points across stereo image pairs.}
	\label{tab:dic_par}
	\begin{tabular}{l c}
        \hline
        Parameter \hspace{25mm} & Specification \\
		\hline
		Formulation \hspace{20mm} & Backward subtractive \\
		Shape functions & Parabolic ($2^{\text{nd}}$ order)\\
		Intensity interpolant & $5^{\text{th}}$ order B-Spline approximant \\
		Subset size & $31\times31$ \si{\pixel} \\
		ROI sampling & $5\times5$ \si{\pixel} or $4\times4$ \si{\pixel} regular grid \\
		Reference updating & Never \\
		Image preprocessing & None\\
		RANSAC control & Active \\
		\hline
	\end{tabular}
\end{table}

Three-dimensional shape reconstruction was achieved by the MATLAB {\texttt{triangulate}} function, which return the positions of points in space by triangulating matched points across the two cameras. The image coordinates of the corresponding points were corrected for lens distortion using the MATLAB {\texttt{undistortPoints}} function and the intrinsic calibration parameters of the two cameras. In particular, the coordinates of each point are defined in a reference frame attached to the master camera (the left one in our setup), with the $x$ and $y$ axes conventionally oriented as in the digital images, and the $z$ axis pointing towards the world reference frame.
The accuracy of the reconstruction process was assessed on the calibration target, by comparing the nominal and measured distances between pairs of circles. Measurements on \num{189} points yielded a mean error of about \num{0.12}\% with a standard deviation of \num{0.16}\%.
Since the PEH samples were imaged while immersed in the electrolyte solution, we also assessed the effect of refraction at the air/electrolyte solution interface on the reconstruction process. To this end, images of the calibration target were acquired under the same conditions used for the hydrogel samples. The comparison between measurements performed under the two conditions (in air or in electrolyte solution) yielded a mean relative error of \num{0.69}\% with a standard deviation of \num{0.04}\%, 
indicating that the effect of the air/electrolyte solution interface on the reconstruction process was negligible.
The identification of matching points across image pairs was performed by means of 2D-DIC~\citep{Sutton2009-DIC, Baldi2013-RSiDIC, BaldiOLEN2022, Baldi_2024} analysis using the research software ABc\footnote{The software ABc is developed and maintained by Prof.~Antonio Baldi from the Universit\`{a} degli Studi di Cagliari, Italy, and is available by contacting Prof.~Antonio Baldi at \href{mailto:antonio.baldi@unica.it}{antonio.baldi@unica.it}.}, which provides both local and global analysis approaches. In the present study, we exploited the local, or subset-based, approach, with the software parameters reported in Table~\ref{tab:dic_par}. These were carefully selected to account for both the speckle pattern density and the light conditions.

The 3D-DIC analysis used a region of interest encompassing the entire sample surface. For subsequent analyses, a right-handed laboratory reference frame was defined  with $\mathbf{e}_3$ normal to the bottom surface of the electrolytic cell and $\mathbf{e}_1$ aligned with the electric field $\mathbf{E}$.
Finally, the reconstructed points were fitted with a bivariate thin-plate smoothing spline, which includes a regularization term penalizing the integrated squared norm of the surface curvature. In particular, we used the MATLAB function \texttt{tpaps} with the smoothing parameter, which controls the relative importance of data fidelity and smoothness, set to $p = 0.05$.

In preparation for each experimental trial, the electrolytic cell was filled with \SI{60}{\milli\liter} of electrolyte solution, and the microscope focal plane adjusted approximately \SI{4.7}{\milli\meter} above the cell bottom surface to prevent the PEH sample from contacting it during stereo imaging. Before application of the electric potential across the electrodes, the elliptic hydrogel sample was gently placed at the center of the electrolytic cell and oriented with its major axis along the normal to the electrodes, while ensuring that it remained in a planar configuration. Upon application of a constant potential difference at the electrodes, a uniform electric field was generated in the electrolyte solution and the sample motion recorded at a frame rate of \num{150}\,fps through the two digital cameras, which were synchronized via a hardwired trigger signal generated by an Arduino UNO R3 board.

\subsection{Experimental results}
\label{sec:exp_res}

Here we report the experimental results on electric-field actuated PEH plates, fabricated and tested as described in the previous sections. The active response of the hydrogel samples was explored for a potential difference at the electrodes ranging between \SI{20}{\volt} and \SI{45}{\volt}. A large variety of dynamic responses were observed.
For potential differences up to \SI{35}{\volt}, the samples underwent small deflections from equilibrium, bending around their minor axis towards the cathode, and eventually reaching a stable equilibrium configuration.
This behavior is consistent with the observations reported by~\citet{boiardiMinimalActuationControl2024} for slender PEH ribbons. Consistent with the higher energetic cost associated with double-curvature configurations in thin plates, the observed configurations were characterized by curvature in one direction only.
In sharp contrast, above a potential difference of \SI{35}{\volt}, small plate deflections rapidly developed into large-amplitude oscillations, the hallmark of flutter instability.
In this regime, the motion of the sample was characterized by a retrograde wave traveling opposite to the electric field, and propelling the PEH plate through the surrounding fluid with a net displacement towards the cathode.

\begin{figure}[t]
	\centering
		\includegraphics[width=\textwidth]{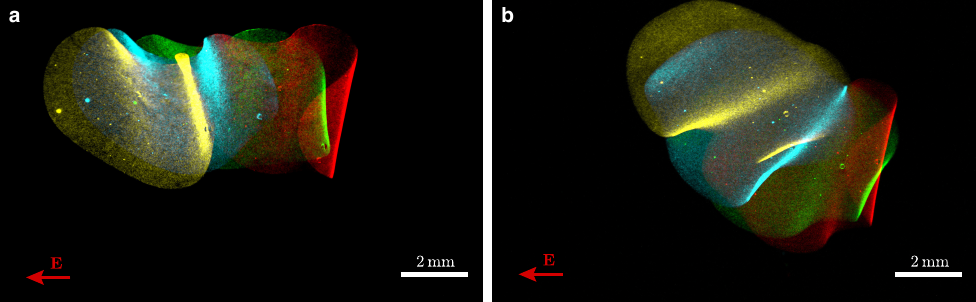}
	\caption{Experimental image sequence of swimming PEH plates as a consequence of flutter instability, induced by a steady and uniform electric field over a time interval of \SI{3.0}{\second}. The colors (red, green, cyan, and yellow) encode the different sampling time of each snapshot, acquired with an offset of \SI{1.0}{\second}. (a) The plate exhibits a straight swimming trajectory over the entire test interval. The electric potential applied at the electrodes is \SI{39}{\volt}. See also Supplementary Video~5. (b) The plate swims along a curved trajectory, drifting away from the straight path. The applied electric potential is \SI{40}{\volt}. See also Supplementary Video~6.} 
	\label{fig:exp_trajectories}
\end{figure}

Across all trials, we observed two qualitatively distinct swimming behaviors, illustrated by the representative cases in Fig.~\ref{fig:exp_trajectories}: an almost straight trajectory, with the direction of motion aligned with the electric field (see Fig.~\ref{fig:exp_trajectories}a and Supplementary Video~5), and a curved trajectory, in which the sample, initially oriented with its major axis along the electric stimulus, progressively deviated from the straight path (see Fig.~\ref{fig:exp_trajectories}b and Supplementary Video~6).
Which of the two behaviors occurred could not be predicted a priori. We attribute this variability to unavoidable differences in the sample configuration at the instant the stimulus was applied.
Interestingly, both behaviors have a qualitative counterpart in the numerical results reported in Fig.~\ref{fig:PlatesNonlinearResults}c and in Supplementary Video~3 for the relevant case of $\lambda = 2/3$.
The complex computational trajectories are indeed  characterized by short intervals of nearly straight swimming and events involving pronounced changes in the direction of motion.   
We interpret the experimental observation in Fig.~\ref{fig:exp_trajectories} as resembling, at least qualitatively, one such short interval of the longer computational trajectories.
The limited size of the electrolytic cell prevents us from observing longer time spans, so that the experiments cannot reproduce the entire swimming behavior predicted by the numerical simulation.

\begin{figure}[p]
	\centering
	\includegraphics[width=\textwidth]{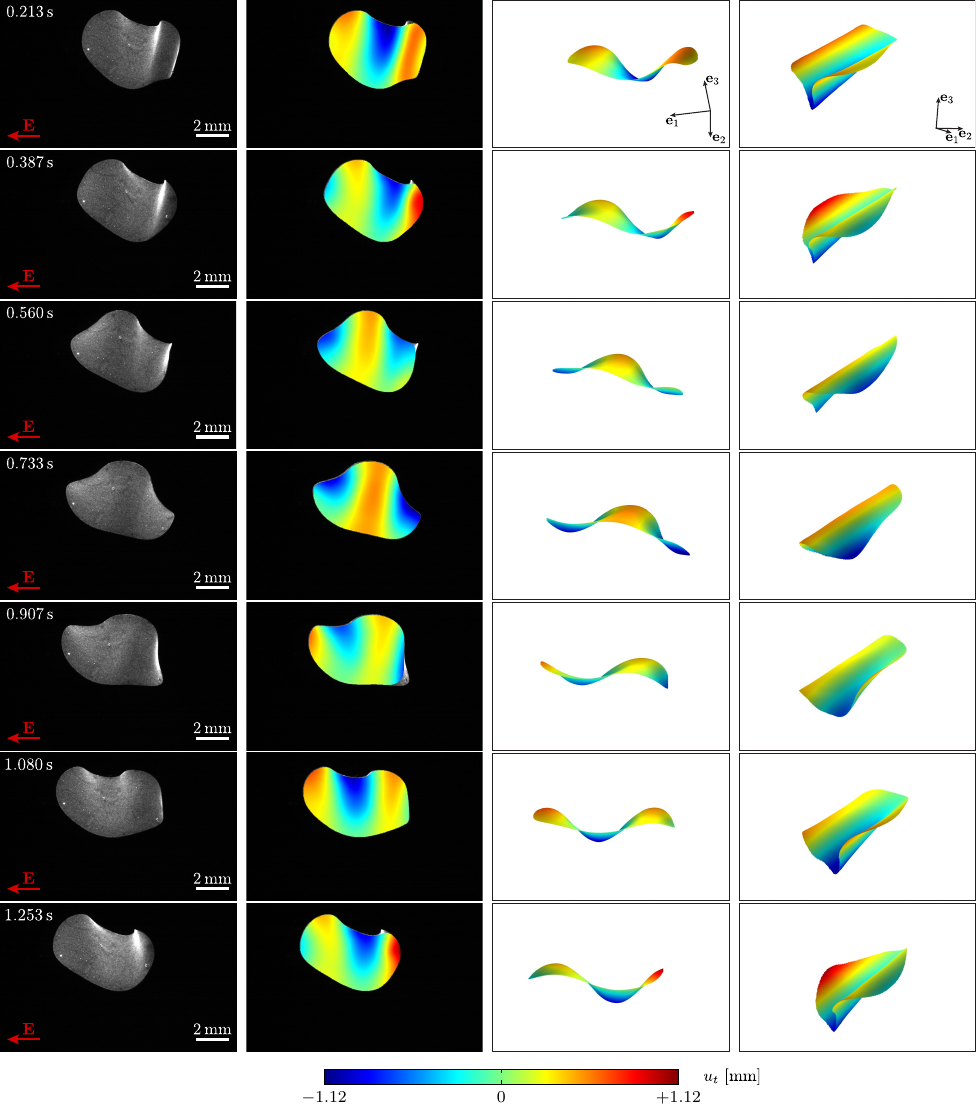}
	\caption{Experimental reconstruction of a swimming PEH plate. The applied voltage is \SI{39}{\volt}. The first column reports the raw images of the sample as it swims towards the cathode. The speckle pattern produced by the beads fluorescence is visible in every image of the sequence; note the saturation around the wave peaks and valleys. The contour plots of the transverse displacement $u_t$ from the midplane of the plate are shown in the second column. The flutter instability produces a retrograde wave that provides a forward thrust to the hydrogel sample. The third and fourth columns report about the side and front views of the experimentally reconstructed plate surface, respectively. The color map again shows the transverse displacement from the midplane, while the two views together fully describe the attitude of the plate during swimming. The triads represent the lab reference frame.
    See also Supplementary Video~5.
	}
    \label{fig:exp_stream_dic}
\end{figure}
Moving from a qualitative to a quantitative characterization, we used 3D-DIC to reconstruct the three-dimensional configuration of the sample shown in Fig.~\ref{fig:exp_trajectories}a at different instants.
The results are reported in Fig.~\ref{fig:exp_stream_dic} for seven successive instants sampled every \num{26} frames, \ie approximately \SI{0.173}{\second} apart, for a total observation window of \SI{1.04}{\second}, see also Supplementary Video~5.

This observation window is slightly longer than the short-time period $T$ of the motion, both as estimated from the experiments and as obtained from the simulations of Section~\ref{sec:plates}.
In the figure, the first two columns illustrate the sample motion as observed by the left camera of the stereo imaging setup. In particular, in the second column, the sample image is color-coded according to the transverse displacement relative to the midplane of the reconstructed shape, thereby highlighting the geometric features that cannot be resolved from raw images alone.
We note the maximum transverse displacement of \SI{1.12}{\milli\meter}, which, when normalized by the size of the ellipse major semi-axis, \ie by \SI{3.51}{\milli\meter}, leads to  $1.12/3.51 \approx 0.32$, in remarkable agreement with maximum dimensionless displacement of \num{0.3} obtained in the numerical simulations.
The third and fourth columns of the figure show the reconstructed shapes from two distinct viewpoints.
These views provide greater detail on the temporal evolution of the sample configuration and reveal the retrograde wave that characterizes it and drives the swimming motion of the sample. All configurations resemble \emph{ruled surfaces}, as observed also in the simulations of~Fig.~\ref{fig:PlatesNonlinearResults}c, with a displacement amplitude that continuously varies across the transverse direction.

%%%%%%%%%%%%%%%%%%%%%%%%%%%%%%%%%%%%%%%%%%%%%%%%%%%%%%%%%%%%%%%%

\section{Conclusions}
\label{sec:conclusions}

In this work, we have investigated the dynamic morphing of PEH shells subjected to constant and uniform electric fields. From a modeling perspective, this was achieved by extending to relevant geometries the description of the active response to electric fields of PEH rods~\citep{cicconofriNonreciprocalOscillationsPolyelectrolyte2023,boiardiMinimalActuationControl2024,boiardiBreakingLeftrightSymmetry2024,boiardi2026three} and embedding it within the framework of non-Euclidean morphoelastic shells.
In carrying out the study, particular emphasis was placed on exploring flutter instability as a mechanism to trigger sustained motion from the interplay between active reconfiguration, shell mechanics, and viscous fluid interactions.

Two prototypical geometries were examined in depth through linear stability analyses and finite element simulations: elliptic plates subjected to an in-plane electric field and cylindrical shells subjected to an electric field directed along their axis.
In both case studies, we identified the parameter ranges over which flutter instability occurs and found that, beyond its onset, the instability develops into sustained nonreciprocal motions in the nonlinear regime.

For the case of elliptic plates, the mechanical problem is purely flexural in the linear regime, while the competition between bending and stretching energies becomes crucial in determining the system behavior at large deformations. The emerging shapes exhibit nearly vanishing Gaussian curvature and give rise to complex swimming trajectories whose characteristics depend sensitively on the plate aspect ratio.
In sharp contrast, for the case of cylindrical shells the membrane and flexural problems are coupled in the linear regime. In this regard, it was demonstrated that the interplay between these deformation modes is essential to sustain flutter in curved geometries, as unstrechable shells cannot undergo this form of instability.
Extending the analysis to the nonlinear regime, it was shown that cylindrical shells develop nonreciprocal traveling waves along their axis, characterized by circumferential lobes whose number is predicted through a semi-analytical dimensional reduction of the mechanical problem. These dynamic shape reconfigurations are reminiscent of peristaltic waves and generate a net thrust in the surrounding fluid, thereby broadening the range of actuation mechanisms achievable in PEH-based active systems.

The study was not limited to theoretical investigations but also included experimental activities carried out on millimeter-sized elliptic plates fabricated from PAAm-co-SA hydrogel by means of UV photolithography. 
The dynamic response of this model system to electric stimuli of different intensities was thoroughly investigated, providing evidence of untethered actuation of such PEH structures through the exploitation of flutter instability. The experimental activities included the application of 3D-DIC techniques for the reconstruction of the three-dimensional sample geometry and supported the numerical findings concerning both the swimming dynamics of the plates and the geometric nature of their deformed configurations, thereby providing experimental validation of the predictive capabilities of the developed morphoelastic model.

The present study focused on two prototypical test cases, paving the way for untethered soft actuators operating under simple, unmodulated stimuli: small-scale swimming devices in the case of plates and valveless pumps or mixers in the case of cylindrical shells.
We emphasize that the geometries here considered represent only a limited subset of the configurations encompassed by the model.
The theoretical framework could be applied for the analysis of other systems, with spherical, toroidal, and axisymmetric shells representing natural candidates, each potentially exhibiting a broad repertoire of functional dynamic responses.
In addition, while particular attention has  been devoted to exploring the potential of flutter, this is not the only form of instability that the investigated systems may support. From a general standpoint, the development of spontaneous curvature in PEH shells in response to electric stimuli gives rise to geometric frustration, which in turn drives instability mechanisms such as buckling and snapping in thin structures~\citep{pezzullaCurvatureInducedInstabilitiesShells2018,giudiciCurvaturedrivenInstabilitiesThin2022}. A distinctive feature of PEH shells is that geometric incompatibility is not permanently built into the system, but is actively generated by the applied external stimulus. This offers the possibility of triggering and controlling instability-mediated dynamic responses, and potentially reversing them, through the electric field alone~\citep{liSpontaneousRapidElectroactuated2022}.
Pursuing this line of investigation would require, on the one hand, the development of suitable experimental protocols for the synthesis and testing of such complex systems and, on the other hand, the development of refined hydrodynamic coupling models to accurately predict their fluid-structure interactions and resulting locomotion or fluid-transport performance.

\section*{Acknowledgements}

The authors acknowledge financial support from the Italian Ministry of University and Research (MUR) through the grant `Dipartimenti di Eccellenza 2023-2027 (Mathematics Area)' and from the European Union\,--\,Next Generation EU in the framework of the PRIN~2022 project no.~2022NNTZNM \lq Unveiling embodied intelligence in natural systems for bioinspired actuator design' (CUP:\,G53D23001240006). The authors thank Prof.~Antonio Baldi for providing access to the ABc software used for the DIC analyses carried out in the present study. The authors are members of the \lq Gruppo Nazionale di Fisica Matematica' (GNFM) of the `Istituto Nazionale di Alta Matematica' (INdAM).

\section*{Description of Supplementary Videos}

All Supplementary Videos are available through the SISSA Open Data repository at \url{https://hdl.handle.net/20.500.12928/OFBHKX}.
All finite element simulations reported in the Supplementary Videos~1\,--\,4 were performed for $\nu = 0.3$. In these Supplementary Videos, $t$ denotes dimensionless time. In particular:
\begin{itemize}
	\item Supplementary Video~1 shows the motion and the straight swimming trajectory of a ribbon-like plate ($\lambda = 1/10$) for $\chi = 36.768$, $\xi = 2.967 \times 10^{-8}$, and $\eta = 8.171 \times 10^{-3}$. The time evolution of the transverse displacement of the plate center is shown at the bottom.
	\item Supplementary Video~2 shows the motion of a circular plate ($\lambda = 1$) for $\chi = 36.768$, $\xi = 2.967 \times 10^{-8}$, and $\eta = 8.171 \times 10^{-3}$. After a transient phase, the plate follows a trajectory at an angle with the external stimulus while rotating about the vertical axis. The time evolution of the transverse displacement of the plate center and of the points located at $(0,\lambda)$ and $(0,-\lambda)$ in the parametric domain $\Omega$ is shown at the bottom in green, red, and blue, respectively.
	\item Supplementary Video~3 shows the motion and the curved, swing-like swimming trajectory of an elliptic plate ($\lambda = 2/3$) for $\chi = 36.768$, $\xi = 2.967 \times 10^{-8}$, and $\eta = 8.171 \times 10^{-3}$. The time evolution of the transverse displacement of the plate center is shown at the bottom, reflecting the complex dynamics of the system.
	\item Supplementary Video~4 shows the periodic, nonreciprocal shape changes of two cylindrical shells, both realized for $\eta = 1.0 \times 10^{-3}$. For the system reported on the left, $\chi = 700$, $\lambda = 0.2$, and $\xi = \xi_\mathrm{ref}$, for which the linear stability analysis predicts a leading circumferential eigenmode with $n_\mathrm{max}=2$. For the system reported on the left, $\chi = 800$, $\lambda = 0.27$, and $\xi = 5\, \xi_\mathrm{ref}$, for which the linear stability analysis predicts a leading circumferential eigenmode with $n_\mathrm{max}=3$. Note that the nonlinear, large-amplitude  oscillations resemble those predicted from the linear analysis.
\end{itemize}

Supplementary Videos~5 and~6 show recordings of experiments conducted on PEH elliptic plates of aspect ratio $\lambda = 2/3$ swimming in the surrounding fluid under a steady and uniform electric field. In particular:
\begin{itemize}
	\item Supplementary Video~5 shows an elliptic plate   exhibiting a straight swimming trajectory towards the cathode over the entire test interval. The contour plots depict the transverse displacement $u_t$ from the midplane at different instants, as reconstructed through 3D-DIC. The electrical potential applied to the electrodes is \SI{39}{\volt}.
	\item Supplementary Video~6 shows an elliptic plate swimming towards the cathode along a curved trajectory, drifting away from the straight path. The applied electrical potential is \SI{40}{\volt}.
\end{itemize}
%

%%%%%%%%%%%%%%%%%%%%%%%%%%%%%%%%%%%%%%%%%%%%%%%%%%%%%%%%%%%%%%%%

%\newpage
\appendix

\section{Estimate of the resistive coefficients}

\subsection{Elliptic plates}
\label{sec:appViscPlate}
The resistive coefficients $\zeta_i$ in~\eqref{eq:viscDragNonDim} for an elliptic plate are estimated from those of an ellipsoidal body moving in a viscous fluid at low Reynolds number~\citep{perrinMouvementBrownienDun1934, happelLowReynoldsNumber1991, montenegro-johnsonMicroscaleFlowDynamics2017}.
Consider such a body with semi-axes $\{A,B,C\}$, oriented along the frame $\{\mathbf{e}_1,\mathbf{e}_2,\mathbf{e}_3\}$ and ordered, without loss of generality, as $C \leq B \leq A$. Assume the body translates with velocity $\mathbf{w} = w_A \, \mathbf{e}_1 + w_B \, \mathbf{e}_2 + w_C \, \mathbf{e}_3$ in an unbounded fluid region of viscosity $\mu$.
The fluid exerts on the ellipsoid a viscous drag  $ \mathbf{F} = -\mu_A \, w_A \, \mathbf{e}_1  - \mu_B \, w_B \, \mathbf{e}_2 - \mu_C \, w_C \, \mathbf{e}_3$, which is determined by solving the corresponding boundary-value Stokes problem.
Upon introducing the elliptic integrals
\begin{align}
    \label{eq:appEllInt1}
    \phi &= \int_0^\infty \frac{\mathrm{d}s}{\sqrt{(A^2+s) (B^2+s) (C^2+s)}} , \\
    \label{eq:appEllInt2}
    \psi_1 &= \int_0^\infty \frac{\mathrm{d}s}{(A^2+s) \sqrt{(A^2+s) (B^2+s) (C^2+s)}} , \\
    \label{eq:appEllInt3}
    \psi_2 &= \int_0^\infty \frac{\mathrm{d}s}{(B^2+s) \sqrt{(A^2+s) (B^2+s) (C^2+s)}} , \\
    \label{eq:appEllInt4}
    \psi_3 &= \int_0^\infty \frac{\mathrm{d}s}{(C^2+s) \sqrt{(A^2+s) (B^2+s) (C^2+s)}} ,
\end{align}
the resistive coefficients for the ellipsoid read
\begin{equation} \label{eq:res_coeff_ellipsoid}
    \mu_A = \frac{16 \pi \mu}{\phi + A^2 \psi_1} , \qquad 
    \mu_B = \frac{16 \pi \mu}{\phi + B^2 \psi_2} , \qquad 
    \mu_C = \frac{16 \pi \mu}{\phi + C^2 \psi_3} .
\end{equation}
Modeling the elliptic plate as a flattened ellipsoid with semi-axes $\{A,B,h/2\}$, where $h$ is the plate thickness, a first estimate of the resistive coefficients $\mu_i$ entering~\eqref{eq:viscDrag}  is obtained by distributing the net drag force over the area $\pi A B$ of the ellipse, namely
\begin{equation}
\label{eq:res_coeff_plate}
    \mu_1 = \frac{\mu_A}{\pi A B} , \qquad \mu_2 = \frac{\mu_B}{\pi A B} , \qquad \mu_3 = \frac{\mu_{h/2}}{\pi A B} .
\end{equation}
To obtain the dimensionless resistive coefficients introduced in Section~\ref{sec:nonlinearModel:nonDim}, we rescale the integrals~\eqref{eq:appEllInt1}--\eqref{eq:appEllInt4} as $\phi = \tilde{\phi} / A$ and $\psi_i = \tilde{\psi}_i / A^3$, where
\begin{align} 
    \tilde{\phi} &= \int_0^\infty \frac{\mathrm{d}s}{\sqrt{(1+s) (\lambda^2+s) [(\eta/2)^2+s]}} , \label{eq:NonDimDrag0} \\
    \tilde{\psi}_1 &= \int_0^\infty \frac{\mathrm{d}s}{(1+s) \sqrt{(1+s) (\lambda^2+s) [(](\eta/2)^2+s]}} ,
    \label{eq:NonDimDrag1} \\
    \tilde{\psi}_2 &= \int_0^\infty \frac{\mathrm{d}s}{(\lambda^2+s) \sqrt{(1+s) (\lambda^2+s) [(\eta/2)^2+s]}} ,
    \label{eq:NonDimDrag2} \\
    \tilde{\psi}_3 &= \int_0^\infty \frac{\mathrm{d}s}{[(\eta/2)^2+s] \sqrt{(1+s) (\lambda^2+s) ((\eta/2)^2+s)}} ,
    \label{eq:NonDimDrag3}
\end{align}
are the corresponding integrals in dimensionless form, depending only on the geometric groups $\lambda$ and $\eta$. Then, the resistive coefficients~\eqref{eq:res_coeff_plate} take the form $\mu_i = (\mu/B) \, \zeta_i$, with 
\begin{equation}
\label{eq:appEllCoeffNonDim}
    \zeta_1 = \frac{16}{\tilde{\phi} + \tilde{\psi}_1} , \qquad
    \zeta_2 = \frac{16}{\tilde{\phi} + \lambda^2 \tilde{\psi}_2} , \qquad
    \zeta_3 = \frac{16}{\tilde{\phi} + (\eta/2)^2 \tilde{\psi}_3} ,
\end{equation}
functions of $\lambda$ and $\eta$ only, consistently with~\eqref{eq:pi_therom_visc}.

We next further simplify the resistive coefficients under the small-thickness assumption for the plate.
Since $\eta \ll \lambda \approx 1$, the coefficients~\eqref{eq:appEllCoeffNonDim} turn out to be only weakly sensitive to the actual value of $\eta$~\citep{montenegro-johnsonMicroscaleFlowDynamics2017}.
This justifies replacing $\zeta_i$ with their limiting values as $\eta \to 0^+$, with negligible error.
To this end, it is expedient to rewrite the expressions~\eqref{eq:NonDimDrag0}--\eqref{eq:NonDimDrag3} in terms of elliptic integrals.
In what follows we assume $\eta \ll \lambda < 1$, treating the case of a circular plate separately. Following~\cite{byrdHandbookEllipticIntegrals1971}, we obtain
\begin{align}
	\tilde{\phi} &= \frac{2 F(\varphi,k)}{\sqrt{1-(\eta/2)^2}} , \\[1.5mm]
	\tilde{\psi}_1 &= \frac{2 \left[ F(\varphi,k) - E(\varphi,k) \right]}{k^2 [1-(\eta/2)^2]^{3/2}} , \\[1.5mm]
	\tilde{\psi}_2 &= \frac{2 \bigg[ E(\varphi,k) - (1-k^2)F(\varphi,k) - k^2 \sin{\varphi} \cos{\varphi}/\sqrt{1-k^2 \sin^2\varphi} \,\bigg]}{k^2 (1-k^2) [1-(\eta/2)^2]^{3/2}} , \\[1.5mm]
	\tilde{\psi}_3 &= \frac{2 \bigg[ \tan\varphi \sqrt{1-k^2\sin^2\varphi} - E(\varphi,k) \bigg]}{(1-k^2) [1-(\eta/2)^2]^{3/2}} ,
\end{align}
where $F(\varphi,k)$ and $E(\varphi,k)$ are the \emph{incomplete elliptic integrals} of the \emph{first} and \emph{second kind}, respectively, with \emph{amplitude} $\varphi = \arccos\eta$ and \emph{modulus} $k = \sqrt{(1-\lambda^2) / (1-\eta^2)}$.
In the limit of $\eta \to 0^+$ we then obtain

\begin{align}
	\zeta_x (\lambda) := \lim_{\eta \to 0^+} \zeta_1 (\lambda, \eta) =  \zeta_1 (\lambda, 0) &= \frac{8}{K\big(\!\sqrt{1-\lambda^2}\big) + \dfrac{K\big(\!\sqrt{1-\lambda^2}\big) - E\big(\!\sqrt{1-\lambda^2}\big)}{1-\lambda^2}} , \label{eq:LimitResCoeffEll1} \\
	\zeta_y (\lambda) := \lim_{\eta \to 0^+} \zeta_2 (\lambda, \eta) =  \zeta_2 (\lambda, 0) &= \frac{8}{K\big(\!\sqrt{1-\lambda^2}\big) + \dfrac{E\big(\!\sqrt{1-\lambda^2}\big) - \lambda^2 K\big(\!\sqrt{1-\lambda^2}\big)}{1-\lambda^2}} , \label{eq:LimitResCoeffEll2} \\
	\zeta_3 (\lambda) := \lim_{\eta \to 0^+} \zeta_3 (\lambda, \eta) &= \frac{8}{K\big(\!\sqrt{1-\lambda^2}\big)} , \label{eq:LimitResCoeffEll3}
\end{align}
where $K(k) := F(\pi/2,k)$ and $E(k) := E(\pi/2,k)$ are the \emph{complete elliptic integrals} of the \emph{first} and \emph{second kind}, respectively.
The nondimensional resistive coefficients~\eqref{eq:LimitResCoeffEll1}--\eqref{eq:LimitResCoeffEll3} are those used in~\eqref{eq:viscDragNonDim} for an active plate with $0 < \lambda < 1$.
The circular case $\lambda = 1$ must be treated separately and is in fact simpler~\citep{happelLowReynoldsNumber1991}.
Here the integrals~\eqref{eq:NonDimDrag0}--\eqref{eq:NonDimDrag3} can be evaluated in closed form, yielding the limiting nondimensional resistive coefficients
\begin{equation}
\label{eq:LimitResCoeffDisk}
	\zeta_x (1) = \zeta_y (1) = \frac{32}{3 \pi} , \qquad \zeta_3 (1) = \frac{16}{\pi} .
\end{equation}
The same values are recovered as the limit of $\lambda \to 1^-$ of~\eqref{eq:LimitResCoeffEll1}--\eqref{eq:LimitResCoeffEll3}, namely
\begin{equation}
	\lim_{\lambda \to 1^-} \zeta_x (\lambda) = \lim_{\lambda \to 1^-} \zeta_y (\lambda) = \frac{32}{3 \pi} , \qquad
	\lim_{\lambda \to 1^-} \zeta_3 (\lambda) = \zeta_3 (1) = \frac{16}{\pi} .
\end{equation}
For convenience, Table~\ref{tab:appResCoefPlate} lists the numerical values of the dimensionless resistive coefficients $\zeta_i$ for $\lambda \in \{ 1/10, 2/3, 1 \}$, as used in the linear stability analysis and in the nonlinear simulations.
\begin{table}
    \centering
    \footnotesize
    \caption{Dimensionless resistive coefficients $\zeta_i$ for an elliptic plate with $\lambda \in \{ 1/10, 2/3, 1 \}$, as computed from expressions~\eqref{eq:LimitResCoeffEll1}--\eqref{eq:LimitResCoeffDisk}.}
    \begin{tabular}{lccc} 
        \toprule
        Resistive coefficient\hspace{10mm} & \textbf{$\lambda = 1/10$} & \textbf{$\lambda = 2/3$} & \textbf{$\lambda = 1$} \\ 
        \midrule
        \hspace{10mm}$\zeta_x$ & 1.2495  & 2.7100 & 3.3953  \\ 
        \hspace{10mm}$\zeta_y$ & 1.7077 & 2.8979  & 3.3953  \\ 
        \hspace{10mm}$\zeta_3$ & 2.1647  & 4.2011 & 5.0930 \\ 
        \bottomrule
    \end{tabular}
    \label{tab:appResCoefPlate}
\end{table}

\subsection{Cylindrical shells}
\label{sec:appViscCyl}
Unlike ellipsoids, cylindrical shells do not admit an explicit solution of the fluid-structure interaction problem in the Stokes regime. We therefore estimate the local drag coefficients following the same rationale adopted for elliptic plates.
The central modeling assumption is that the hydrodynamic traction exerted by the fluid at a point on the shell midsurface depends only on the local velocity of that point, and is equal to the local drag experienced by a rigid disk of radius equal to that of the cylinder and tangent to the shell at that point.
This amounts to retaining only the dominant local resistance, while neglecting the long-range hydrodynamic interactions between distinct portions of the surface, in analogy with resistive-force theory for slender bodies.

The drag force on a rigid disk translating in a Stokes flow is known in closed form, as the degenerate limit of the results~\eqref{eq:res_coeff_ellipsoid} for an ellipsoid results. In particular, a disk of radius $R$ translating edgewise with velocity $U$ in a fluid of viscosity $\mu$ experiences a drag force of magnitude $(32/3)\,\mu R U$, whereas broadside translation of the disk at the same velocity gives rise to a drag force of magnitude $16\,\mu R U$, in both cases opposing the direction of motion.
Since the disk is assumed tangent to the shell surface, its plane is spanned by $\mathring{\mathbf{a}}_s$ and $\mathring{\mathbf{a}}_z$, while its normal coincides with $\mathring{\mathbf{a}}_3$. Consequently, the circumferential and axial components of the shell velocity determine the edgewise drag, whereas the radial component of the shell velocity determines the broadside drag.
We estimate the local drag coefficients for the shell by distributing the net drag force acting on the disk over its area $\pi R^2$, obtaining
\begin{equation}
    \mu_z = \frac{32}{3 \pi} \frac{\mu}{R} =: \frac{\mu}{R} \zeta_z , \qquad
    \mu_s = \frac{32}{3 \pi} \frac{\mu}{R} =: \frac{\mu}{R} \zeta_s , \qquad \mu_3 = \frac{16}{\pi} \frac{\mu}{R} =: \frac{\mu}{R} \zeta_3 .
\end{equation}

In contrast to the case of elliptic plates, for which the resistive coefficients depend on the aspect ratio $\lambda$, the dimensionless coefficients $\zeta_s = \zeta_z = 32/(3\pi)$ and $\zeta_3 = 16/\pi$ are independent of the geometry of the cylindrical shell. These values coincide with the resistive coefficients of a circular plate in the vanishing-thickness limit~\eqref{eq:LimitResCoeffDisk}, providing a consistency check on our computations. In particular, the resistance to radial motion, which displaces the surrounding fluid broadside, is larger than the resistance to tangential motion, as expected on physical grounds.
We stress that a complete and accurate description of the fluid-shell interaction lies beyond the scope of the present work. Here, viscous forces are included to qualitatively assess their influence on the dynamic shape reconfigurations of interest, whose main features are not expected to depend sensitively on the details of the drag model. A refined hydrodynamic description accounting for the nonlocal interactions neglected here is left for future work.
%

%%%%%%%%%%%%%%%%%%%%%%%%%%%%%%%%%%%%%%%%%%%%%%%%%%%%%%%%%%

\section{Parameters choice and estimation}
\label{sec:appElasticPEH}
We collect here the details on the choice of the parameters used in the stability analyses and in the nonlinear simulations of Sections~\ref{sec:plates} and~\ref{sec:cylindricalShells}.
As established in Section~\ref{sec:nonlinearModel:nonDim}, the response of the model is governed by the physical parameters only through the dimensionless groups $\{\nu, \lambda, \eta, \xi, \chi\}$, which are gathered in Table~\ref{tab:adm_par}.
Consequently, the physical parameters affect the behavior of the system only via the combinations by which they enter the dimensionless groups.
Guided by this observation, and considering that a complete experimental characterization of the model parameters lies beyond the scope of the present work, we content ourselves with representative estimates.

Within our theory, the Young modulus $Y$ and the Poisson ratio $\nu$ are the material parameters characterizing the isotropically swollen hydrogel in the small-strain regime. We set the Poisson ratio to $\nu = 0.3$, following~\citet{yoonPoroelasticSwellingKinetics2010} and~\citet{bouklasSwellingKineticsPolymer2012}, where the material parameters of similar hydrogels were characterized.
As for the Young modulus $Y$ and the relaxation time $\tau$, these are informed by the estimates of~\citet{boiardiPhDThesis2025}, reporting for similar PEH a Young modulus of about \SI{37.5}{\kilo\pascal} and a relaxation time of about \SI{0.33}{\second}.
Since the fabrication protocols relevant to our study differ from those of the reported literature, we adopt the representative values of $Y = \SI{30}{\kilo\pascal}$ and $\tau = \SI{1}{\second}$, with the understanding that they capture the correct order of magnitude rather than providing a precise material characterization.
Regarding the dynamic viscosity of the electrolyte solution, this is taken to be that of water at \SI{25}{\celsius}, namely we set $\mu = \SI{0.89}{\milli\pascal\second}$.
Together, these choices yield the value of $\xi_\mathrm{ref} = \mu / (Y \tau) = 2.967 \times 10^{-8}$.
This is the reference value of $\xi$ used in the numerical simulations of Section~\ref{sec:plates:simulations}. Recall that in Section~\ref{sec:cylindricalShells:stability} we explored  the stability of the system for $\xi \in \{\xi_\mathrm{ref}, 5\,\xi_\mathrm{ref}, 20\,\xi_\mathrm{ref}\}$, so that the qualitative conclusions are robust with respect to the specific choice of $Y$ and $\tau$. Equivalently, a variation of $\xi$ within this range can be interpreted physically as a change in the surrounding fluid (a higher dynamic viscosity) or in the hydrogel (a lower stiffness or a shorter relaxation time).

We now turn to the geometric parameters. In Section~\ref{sec:plates}, those of the plates fabricated in Section~\ref{sec:experiments} are adopted, for consistency between numerics and experiments.
At the fabrication stage, the plate axes measured \SI{4.92}{\milli\meter} and \SI{3.32}{\milli\meter}, while the plate thickness was approximately \SI{20}{\micro\meter}, as determined by the thickness of the spacers. Once swollen in the electrolyte solution, the axes increased to \SI{7.02}{\milli\meter} and \SI{4.75}{\milli\meter}, corresponding to a swelling factor of \num{1.43} and hence to an estimated swollen thickness of \SI{28.6}{\micro\meter}.
Consistently with this fabrication, in our simulations we identify the characteristic length $L_1$ with the major semi-axis of the plate and set $L_1 = \SI{3.5}{\milli\meter}$ and $h = \SI{28.6}{\micro\meter}$, giving $\eta = h / L_1 = 8.1714 \times 10^{-3}$ and hence $\xi_\mathrm{ref} / \eta^3 = 0.054$.
The loading parameter $\chi = 36.768$ is chosen so that the system is in the flutter regime for all the considered values $\lambda = \{1/10, 2/3, 1\}$. In particular, $\lambda = 2/3$ matches the aspect ratio of the plates used in the experiments of Section~\ref{sec:experiments}.
Since the analysis of Section~\ref{sec:cylindricalShells} lacks an experimental counterpart, we fix $\eta = 1.0 \times 10^{-3}$ as a reasonable ratio between the thickness of the cylindrical shell and its length, and we explore the stability of the system over the range of $\xi$ introduced above. This amounts to studying the same class of material in a different geometry, while acknowledging, through the variation of $\xi$, the uncertainty inherent in the parameter estimates.

%%%%%%%%%%%%%%%%%%%%%%%%%%%%%%%%%%%%%%%%%%%%%%%%%%%%%%%%%%

\section{Numerical implementation of the stability analyses}

We collect here the details of the finite element discretizations adopted for the linear stability analyses and of the strategy used to compute the stability maps over the parameter space.

\subsection{Elliptic plates}
\label{sec:appDetailsPlatesStab}

The eigenvalue problem stemming from~\eqref{eq:Balance3PlateLin}--\eqref{eq:EVLAW_WEAK} upon accounting for the time-harmonic assumptions~\eqref{eq:TH_PLATE} was solved numerically via the finite element method, using the commercial software COMSOL Multiphysics 6.3. The problem was discretized in space using the quintic Argyris finite element for $\hat{u}_3$, which ensures $\mathcal{C}^1$-continuity of the transverse displacement, and the quadratic Lagrange finite element for $\hat{\kappa}$. Upon denoting the unknown vector by $\hat{\mathbf{X}} = (\hat{u}_{3}, \hat{\kappa})^\mathsf{T}$, the discretized eigenvalue problem reduces to
\begin{equation}
\label{eq:LinEigPlate_num}
    \omega \, \hat{\mathbf{X}}_N = \mathbb{K}_N \, \hat{\mathbf{X}}_N ,
\end{equation}
where $\hat{\mathbf{X}}_N$ is the vector of nodal unknowns and $\mathbb{K}_N$ is the discrete system matrix. The elliptic parametric domain $\Omega$ was discretized into triangular mesh elements, with a mesh symmetric with respect to the $x$ and $y$ axes.
To ensure numerical stability across the parameter space, an adequate number of triangular mesh elements was determined to be 328 for $\lambda = 1/10$, 848 for $\lambda = 2/3$, and 748 for $\lambda = 1$, corresponding to 2499, 5807, and 5139 degrees of freedom, respectively.
Computing the whole set of eigenvalues and corresponding eigenvectors over a fine parameter grid $(\xi/\eta^3, \chi)$ to generate high-resolution stability maps was computationally prohibitive.
For this reason, at each value of $\lambda$, we solved the eigenvalue problem at the point $(\xi/\eta^3, \chi) = (0.5 , 50)$ using the LAPACK algorithm, computing all eigenvalues and identifying the unstable ones, \ie those with positive real parts.
Subsequently, these unstable eigenvalues were tracked through the region $(\xi/\eta^3, \chi) \in [0.002, 0.5] \times [0, 50]$ in $250 \times 251$ steps using the ARPACK algorithm, which reduces the computational cost once an initial guess is provided. This procedure led to the identification of the transitions between blue, yellow, and green regions in the first row of Fig.~\ref{fig:PlateGeometry}b, while the eigenvalues and the corresponding eigenmodes in the second row of the same figure were obtained directly via the LAPACK algorithm for $(\xi/\eta^3, \chi) = (0.054, 36.768)$.

\subsection{Cylindrical shells}
\label{sec:appDetailsCylindersStab}
The eigenvalue problem obtained by combining~\eqref{eq:LinCyl1DtimeHarmonic} with the system~\eqref{eq:Balance1CylLin1D}--\eqref{eq:EvLawCylLin1D} was solved numerically by the finite element method, using the commercial software COMSOL Multiphysics 6.3. The problem was discretized in space with cubic Hermite elements for $\hat{b}_s^n$, $\hat{a}_z^n$, and $\hat{a}_3^n$, and quadratic Lagrange elements for $\hat{a}_\kappa^n$. Collecting the unknowns in the vector $\hat{\mathbf{X}}^n = ( \hat{b}_s^n, \hat{a}_z^n, \hat{a}_3^n, \hat{a}_\kappa^n )^\mathsf{T}$, the discretized eigenvalue problem reads
\begin{equation}
\label{eq:LinEigCyil_num}
    \omega_n \, \hat{\mathbf{X}}_{N}^n = \mathbb{K}_{N}^n \, \hat{\mathbf{X}}_{N}^n ,
\end{equation}
where $\hat{\mathbf{X}}_{N}^n$ is the vector of nodal unknowns associated with the circumferential mode $n$, and $\mathbb{K}_{N}^n$ is the corresponding discrete system matrix. The domain $[0,1]$ was discretized into $50$ non-uniform elements, geometrically graded with growth factor $\exp(0.04)$ so as to cluster the finest elements near the edge $z=0$. This refinement was required for numerical accuracy, as the waves traveling along the cylinder axis give rise to a boundary layer at $z=0$~\citep{efratiBucklingTransitionBoundary2009}, which must be adequately resolved. The resulting discretization amounts to $407$ degrees of freedom for each decoupled problem indexed by $n$.
The dimensional reduction allowed for the computation of the entire eigenvalue spectrum for any desired $n$. In practice, we fixed $M=20$ and solved the problem for $n=0,\dots,M$.
For each value of $\xi \in \{ \xi_\mathrm{ref}, 5 \, \xi_\mathrm{ref}, 20 \, \xi_\mathrm{ref} \}$, the eigenvalue problems were solved over the region $(\lambda, \chi) \in [0.05, 0.5] \times [0, 1400]$, sampled on a $91 \times 140$ grid, see Fig.~\ref{fig:CylGeometry}b. At each grid point, the full set of eigenvalues was computed for every $n$ using the LAPACK algorithm. Then, $\omega_\mathrm{max}$ was identified as the eigenvalue with the largest real part, and $n_\mathrm{max}$ as the corresponding circumferential mode.

%%%%%%%%%%%%%%%%%%%%%%%%%%%%%%%%%%%%%%%%%%%%%%%%%%%%%%%%%%

\section{Numerical implementation of the nonlinear models}

Numerical simulations were performed via the finite element method using the commercial software COMSOL Multiphysics 6.3. The problem was semi-discretized in space using the quintic Argyris finite element for the three components $u_i$ of the displacement, ensuring $\mathcal{C}^1$-continuity~\citep{haleSimpleExtensiblePlate2018a}, and the quadratic Lagrange finite element for the spontaneous curvature $\kappa$. 
Time integration was performed using the implicit generalized-$\alpha$ method, while Newton's iterative method was used to solve the resulting nonlinear algebraic system.

\subsection{Elliptic plates}
\label{sec:appDetailsPlatesNonlin}

For the numerical study reported in Section~\ref{sec:plates:simulations}, the weak form~\eqref{eq:PVWExpNonDim} was tested with $\mathbf{v} = v_x \,\mathbf{e}_1$, $\mathbf{v} = v_y \, \mathbf{e}_2$, and $\mathbf{v} = v_3 \, \mathbf{e}_3$ to reduce it to three scalar weak equations, coupled with~\eqref{eq:evLawNonDim} recast in weak form as
\begin{equation}
\label{eq:evLawNonDimWeak}
    \int_\Omega \bigg\{\Big[ \partial_{t} \kappa + \kappa + \chi \, \mathbf{e}_1 \cdot \mathbf{a}_3 \Big]  \psi\bigg\} \, \mathrm{d}x \mathrm{d}y = 0 , \qquad \forall \psi .
\end{equation}

The system of weak equations was solved imposing, at $t=0$, zero spontaneous curvature and a perturbation of the trivial equilibrium with a displacement field corresponding to a rotation of the plate about $\mathbf{e}_2$ by an angle $\pi / 512$.
The elliptic parametric domain $\Omega$ was discretized into triangular mesh elements, with a mesh symmetric with respect to the $x$ and $y$ axes and corresponding, for consistency, to that used in the linear stability analysis. The number of triangular mesh elements was 328 for $\lambda = 1/10$, 848 for $\lambda = 2/3$, and 748 for $\lambda = 1$, corresponding to 6069, 13899, and 12303 spatial degrees of freedom, respectively.
Time integration was performed with a time step $\Delta t = 0.002$, saving the results every five time steps to optimize memory usage.

\subsection{Cylindrical shells}
\label{sec:appDetailsCylindersNonlin}

For the numerical study reported in Section~\ref{sec:cylindricalShells:simulations}, the weak form~\eqref{eq:PVWExpNonDim} was tested with $\mathbf{v} = v_s \, \mathring{\mathbf{a}}_s$, $\mathbf{v} = v_z \, \mathring{\mathbf{a}}_z$, and $\mathbf{v} = v_3 \, \mathring{\mathbf{a}}_3$, respectively, to reduce it to three scalar weak equations, coupled with~\eqref{eq:evLawNonDim} recast in weak form as
\begin{equation}
\label{eq:evLawNonDimWeakCyl}
    \int_\Omega \bigg\{\Big[ \partial_{t} \kappa + \kappa + \chi \, \mathbf{e}_3 \cdot \mathbf{a}_z \Big]\bigg\} \psi \, \mathrm{d}s \mathrm{d}z = 0 , \qquad \forall \psi .
\end{equation}
As required by the weak form of the equations, periodicity in $s=0$ and $s=2\pi\lambda$ was guaranteed by enforcing
\begin{equation}
    u_i (0,z,t) = u_i(2 \pi \lambda,z,t) , \quad
    \kappa (0,z,t) = \kappa(2 \pi \lambda,z,t) , \quad
    \partial_\alpha u_i (0,z,t) = \partial_\alpha u_i(2 \pi \lambda,z,t) , \quad i = \{s,z,3\}, ~~~ \alpha = \{s,z\} ,
\end{equation}
for any $z \in [0,1]$ and $t \geq 0$. We remark that 
COMSOL Multiphysics allows the implementation of such conditions via the \texttt{Linear Extrusion} coupling operator.
The system of weak equations was solved assuming, at $t=0$, a spontaneous curvature $\kappa(s,z,0) = 0.1$ in the whole domain, as a small perturbation from the trivial equilibrium, and zero initial displacement. 
The rectangular parametric domain $\Omega$ was discretized into a structured, non-uniform grid of $50 \times 16$ rectangular cells. The 50 subdivisions along the $z$ direction were geometrically graded with growth factor $\exp(0.04)$, clustering the finest elements near one boundary for numerical accuracy. 
In fact, the waves traveling along the cylinder axis are responsible for a boundary layer~\citep{efratiBucklingTransitionBoundary2009} on the edge $z=0$, so that a local mesh refinement is required at the boundary.
Each cell was then split into four triangular elements by inserting a central node connected to its corners, resulting in 3200 triangular mesh elements, corresponding to a total of 51137 spatial degrees of freedom.
Time integration was performed with a time step $\Delta t = 0.01$.

%%%%%%%%%%%%%%%%%%%%%%%%%%%%%%%%%%%%%%%%%%%%%%%%%%%%%%%%%%

\bibliographystyle{elsarticle-harv}
\bibliography{bibliography}

\end{document}